\documentclass{IEEEtran}
\usepackage{cite}
\usepackage{amsmath,amssymb,amsfonts}
\usepackage{algorithmic}
\usepackage{graphicx}
\usepackage{textcomp}
\usepackage{multirow}
\usepackage{caption}
\usepackage{array}
\usepackage{subcaption}
\usepackage{url}
\usepackage{hyperref}
\usepackage{textcomp}
\usepackage{xcolor}
\usepackage{soul}
\sethlcolor{yellow}
\usepackage{float}

\newcolumntype{M}[1]{>{\centering\arraybackslash}m{#1}}
\newcolumntype{L}[1]{>{\raggedright\arraybackslash}m{#1}}

\def\BibTeX{{\rm B\kern-.05em{\sc i\kern-.025em b}\kern-.08em
    T\kern-.1667em\lower.7ex\hbox{E}\kern-.125emX}}
\begin{document}
\title{On TinyML and Cybersecurity: Electric Vehicle Charging Infrastructure Use Case}
\author{Fatemeh Dehrouyeh\thanks{Fatemeh Dehrouyeh: Electrical and Computer Engineering, University of Western Ontario, London, Ontario, Canada. Email: fdehrouy@uwo.ca}, 
        Li Yang\thanks{Li Yang: Faculty of Business and Information Technology, Ontario Tech University, Oshawa, Ontario, Canada. Email: li.yang@ontariotechu.ca}, 
        Firouz Badrkhani Ajaei\thanks{Firouz Badrkhani Ajaei: Electrical and Computer Engineering, University of Western Ontario, London, Ontario, Canada. Email: fajaei@uwo.ca}, 
        and Abdallah Shami\thanks{Abdallah Shami: Electrical and Computer Engineering, University of Western Ontario, London, Ontario, Canada. Email: ashami2@uwo.ca}}

\maketitle

\begin{abstract}
As technology advances, the use of Machine Learning (ML) in cybersecurity is becoming increasingly crucial to tackle the growing complexity of cyber threats. While traditional ML models can enhance cybersecurity, their high energy and resource demands limit their applications, leading to the emergence of Tiny Machine Learning (TinyML) as a more suitable solution for resource-constrained environments. TinyML is widely applied in areas such as smart homes, healthcare, and industrial automation.  TinyML focuses on optimizing ML algorithms for small, low-power devices, enabling intelligent data processing directly on edge devices. This paper provides a comprehensive review of common challenges of TinyML techniques, such as power consumption, limited memory, and computational constraints; it also explores potential solutions to these challenges, such as energy harvesting, computational optimization techniques, and transfer learning for privacy preservation. On the other hand, this paper discusses TinyML's applications in advancing cybersecurity for Electric Vehicle Charging Infrastructures (EVCIs) as a representative use case. It presents an experimental case study that enhances cybersecurity in EVCI using TinyML, evaluated against traditional ML in terms of reduced delay and memory usage, with a slight trade-off in accuracy. Additionally, the study includes a practical setup using the ESP32 microcontroller in the PlatformIO environment, which provides a hands-on assessment of TinyML's application in cybersecurity for EVCI.
\end{abstract}


\section{Introduction}
The widespread use of Internet of Things (IoT) technologies has transformed our daily lives, envisioning a world where homes intelligently adjust to preferences~\cite{fang2020study}, and wearable devices maintain effortless connectivity~\cite{dzedzickis2020human}. This seamless integration and communication among devices revolutionize our interaction with the environment.

However, advancements in IoT technology correspond with an escalation in cybersecurity threats. Attackers often take advantage of vulnerable IoT devices as an entry point for compromising critical backend systems, especially in sensitive sectors such as industrial automation, smart grids, transportation, and healthcare services~\cite{stellios2018survey, shaer2023hierarchical}.
This growing concern is further amplified by evolving cybersecurity challenges such as malware, phishing, and social engineering, which are continuously transformed by emerging technologies such as cloud computing, mobile devices, and Artificial Intelligence/Machine Learning (AI/ML)~\cite{jerbi2023beyond}.

Expanding on the discussion of evolving cybersecurity, ML models play a pivotal role in developing advanced systems for malware detection, spam classification, and network intrusion identification. Techniques such as Deep Learning (DL), Support Vector Machines (SVM), and Random Forest (RF) have proven effective in identifying cybersecurity threats~\cite{abdullahi2022detecting, Moubayed_DNS_ML}. These sophisticated models often operate on high-powered servers to identify vulnerabilities and analyze large datasets for unusual patterns or signs of cyber threats~\cite{dutta2021implementation, 9969601}. 

Despite ML models emerging as a strong solution for addressing cyber attacks, depending solely on traditional ML techniques may be inadequate due to their inherent limitations, especially in the context of analyzing IoT data~\cite{YANG2022105366, Manias_Model_Drift}. Traditional ML models often encounter problems related to size, computational complexity, and high energy use, making them poorly suited for effective deployment on resource-limited edge devices commonly employed in IoT systems~\cite{9969601, Cesar_s18113779}. Similarly, current cloud-based ML systems encounter challenges, including high power consumption and issues related to security, privacy, reliability, and response time delays. These challenges highlight the limitations of relying solely on cloud computing for ML tasks~\cite{kallimani2023tinyml, ABUSHARKH201678}.

Deploying ML on IoT devices presents significant challenges, especially regarding battery life. These devices are often placed in locations where regular battery maintenance or replacement is impractical or impossible. This situation raises concerns about cost, environmental impact, and the performance sustainability of such devices, as prolonged battery life becomes crucial for their effective operation. Additionally, the desire for small devices conflicts with the need for large size of batteries~\cite{9174941}. 

Consequently, despite ML advancements in IoT data analysis, real-world deployment of these models encounters significant challenges, notably limited power and storage, especially in devices with constrained batteries~\cite{SABOVIC2023100736, YANG2022105366}. It should be noted that the term limited power refers to the issue of  energy management, including how efficiently a device can operate with the available energy. On the other hand, constrained batteries specifically address the limitations related to the battery's capacity and lifespan, emphasizing the physical and chemical constraints that affect how much energy can be stored and the battery's durability over time.

To address these challenges, a novel concept called TinyML has emerged. The primary objective of TinyML is to design, develop, and deploy optimized ML models on ultra-low-power IoT hardware devices while minimizing energy consumption~\cite{DUTTA2021100461, SABOVIC2023100736}.
TinyML is dedicated to the optimization of ML models while accepting a minor reduction in accuracy as a trade-off for enhanced feasibility and efficiency on edge devices~\cite{10177729}.

TinyML approach presents multiple advantages in devices with limited computational and memory capabilities. Key benefits include decreased response times, minimized bandwidth needs, lower energy usage, and enhanced security and privacy. These improvements are particularly notable in microcontroller-based ML models, which have shown robust performance in various IoT settings. By processing data locally, TinyML not only facilitates real-time threat detection but also strengthens privacy by reducing data transfer through potentially insecure networks~\cite{DUTTA2021100461, RAY20221595, DELNEVO2023100729, 9014355, immonen2022tiny, fi14120363, s20092638, SABOVIC2023100736}.


\begin{table}[h!]
\caption{List of Acronyms}
\centering
\scriptsize
\begin{tabular}{|c|c|}
\hline
\textbf{Acronym} & \textbf{Full Term} \\
\hline
AC & Alternating Current \\ \hline
AI & Artificial Intelligence \\ \hline
AIDS & Anomaly-based Intrusion Detection System \\ \hline
AR & Augmented Reality \\ \hline
AutoML & Automated Machine Learning \\ \hline
BEMS & Building Energy Management Systems \\ \hline
BEV & Battery Electric Vehicle \\ \hline
C\&C & Command and Control \\ \hline
CAN & Controller Area Network \\ \hline
CCLCS & Coordinated Charging Load Control System \\ \hline
CCS & Combined Charging System \\ \hline
CMS & Central Management System \\ \hline
CNN & Convolutional Neural Network \\ \hline
DC & Direct Current \\ \hline
DDoS & Distributed Denial-of-Service \\ \hline
DL & Deep Learning \\ \hline
DNN & Deep Neural Network \\ \hline
DoS & Denial of Service \\ \hline
ELL & Embedded Learning Library \\ \hline
ELM & Edge Learning Machine \\ \hline
EV & Electric Vehicle \\ \hline
EVCS & Electric Vehicle Charging Station \\ \hline
EVSE & Electric Vehicle Supply Equipment \\ \hline
FAR & False Alarm Rate \\ \hline
FDI & False Data Injection\\ \hline
FL & Federated Learning \\ \hline
FTP & File Transfer Protocol \\ \hline
GPU & Graphics Processing Units \\ \hline
HTTP/HTTPS & Hypertext Transfer Protocol Secure \\ \hline
IAT & Inter-Arrival Times \\ \hline
ICMP & Internet Control Message Protocol \\ \hline
IDE & Integrated Development Environment \\ \hline
IDS & Intrusion Detection System \\ \hline
IEC & International Electrotechnical Commission \\ \hline
ISO & International Organization for Standardization \\ \hline
IoT & Internet of Things \\ \hline
MCU & Microcontroller Unit \\ \hline
MitM & Man-in-the-Middle \\ \hline
ML & Machine Learning \\ \hline
MLP & Multi-Layer Perceptron \\ \hline
MMA & Malicious Mode Attack \\ \hline
MQTT & Message Queuing Telemetry Transport \\ \hline
NAS & Neural Architecture Search \\ \hline
NEC & National Electrical Code \\ \hline
NFPA & National Fire Protection Association \\ \hline
NaN & Not a Number \\ \hline
OCPP & Open Charge Point Protocol \\ \hline
PEV & Plug-In Electrical Vehicle \\ \hline
PHEV & Plug-In Hybrid Electric Vehicle \\ \hline
PTQ & post-training quantization \\ \hline
QAT & Quantization-Aware Training \\ \hline
RF & Random Forest \\ \hline
RP & Recommended Practice \\ \hline
RSSI & Radio Signal Strength Indicator \\ \hline
SAE & Society of Automotive Engineers \\ \hline
SIDS & Signature-based Intrusion Detection System \\ \hline
SMBO & Sequential Model-Based Optimization \\ \hline
SPIFFS & Serial Peripheral Interface Flash File System \\ \hline
SSH & Secure Shell \\ \hline
SSL & Secure Socket Layer \\ \hline
STI & Short Time Inoperability \\ \hline
SVM & Support Vector Machine \\ \hline
TCP/IP & Transmission Control Protocol/Internet Protocol \\ \hline
TFLM & TensorFlow Lite Micro \\ \hline
TFLite & TensorFlow Lite \\ \hline
TL & Transfer Learning \\ \hline
TLS & Transport Layer Security \\ \hline
TinyML & Tiny Machine Learning \\ \hline
UDP & User Datagram Protocol \\ \hline
V2G & Vehicle-to-Grid \\ \hline
VSCode & Visual Studio Code \\ \hline
WPT & Wireless Power Transfer \\ \hline
WSL & Windows Subsystem for Linux \\ \hline
XSS & Cross-Site Scripting \\ \hline
\end{tabular}
\label{table:acronyms} 
\end{table}

This paper aims to provide a comprehensive review of TinyML challenges and potential solutions in general cybersecurity contexts. Given that EVCIs are among the most representative and impactful use cases for TinyML~\cite{sarieddine2022investigating, metere2021securing, elkashlan2023intrusion, basnet2021exploring}, we include a detailed case study on EVCI. This case study serves to illustrate how TinyML can be effectively applied to real-world IoT systems, offering a clear understanding of the practical applications and benefits of TinyML in enhancing cybersecurity.

The implementation of TinyML plays a pivotal role in improving cybersecurity in the domain of EVCI~\cite{elkashlan2023intrusion}. EVCIs are susceptible to various cybersecurity vulnerabilities due to their interconnected nature and the sensitive data they handle, such as payment information and real-time energy usage statistics. Integrating the deployment of TinyML algorithms in Electric Vehicles (EVs) aims to establish lightweight yet robust security measures at the edge. This approach focuses on enhancing data privacy, reducing latency, and improving the resilience of critical charging infrastructures against potential cyber threats. To illustrate these benefits, this paper includes a case study on using TinyML in EVCI cybersecurity.

The main contributions of this paper are as follows:
\begin{itemize}
    \item An in-depth exploration of the role and significance of TinyML in enhancing cybersecurity.
    \item A detailed analysis of challenges and constraints in implementing TinyML, with potential solutions.
    \item A comprehensive case study on the application of TinyML for EVCI cybersecurity\footnote{The source code of this paper is available at: \url{https://github.com/Western-OC2-Lab/TinyML_EVCI}}.
    \item An experimental setup and practical implementation of TinyML on the ESP32 microcontroller, including detailed programming practices, communication protocols, and performance insights.
\end{itemize}
The paper is structured as follows: Section II serves as a background, introducing TinyML, discussing various cybersecurity threats, and highlighting TinyML's benefits. Section III delves into the challenges and constraints of using TinyML on devices with limited resources. Section IV elaborates on potential solutions to these challenges, along with key tools and libraries that facilitate TinyML implementation. Section V presents a comprehensive case study on the application of TinyML in EVCIs, focusing on the framework, communication protocols, attack scenarios, and countermeasures. Section VI presents the experimental setup and a detailed analysis of the results. Section VII discusses the practical implementation of TinyML on an ESP32 microcontroller. Section VIII details future research directions. Finally, Section IX concludes the paper by summarizing the key findings and contributions and highlighting the impact of TinyML on IoT and cybersecurity. Additionally, for a comprehensive list of acronyms used in this paper, please refer to Table~\ref{table:acronyms}.

\section{Background: TinyML for Cybersecurity}

This section provides an essential overview of TinyML's role in cybersecurity, beginning with a clear definition of TinyML, its advantages, and practical applications. The focus is then shifted to a critical examination of the cybersecurity domain, where various threats faced by organizations today are detailed. Finally, the importance of TinyML solution in addressing the ongoing challenge posed by cyber threats is emphasized.

\subsection{TinyML Overview}

\subsubsection{Definition of TinyML}

\begin{table*}
\centering
\caption{Advantages of TinyML}
\renewcommand{\arraystretch}{1.25} 
\begin{tabular}{|M{4cm}|L{12cm}|}
\hline
\multicolumn{1}{|c|}{\textbf{Advantage}} & \multicolumn{1}{c|}{\textbf{Description}}
\\
\hline
Bandwidth Efficiency & Integration of ML models in IoT devices reduces the need for transmitting data to the cloud, which results in lower bandwidth requirements~\cite{mi13060851, DUTTA2021100461}. \\
\hline
Latency Reduction & On-device analysis reduces latency and ensures swift decision-making for time-sensitive applications such as healthcare and autonomous vehicles~\cite{fi14120363, 10177729, RAY20221595, DUTTA2021100461}. \\
\hline
Cost Savings & Reduced data traffic and minimal hardware constraints of IoT devices lead to cost savings compared to resource-intensive cloud-based solutions~\cite{10177729, 9166461}. \\
\hline
Energy Efficiency & Local processing on IoT devices consumes significantly less energy compared to wireless data transmission~\cite{9166461, RAY20221595}. \\
\hline
Enhanced Security and Privacy & Minimized data flow reduces the risk of malicious attacks and enhances data security and privacy by design~\cite{fi14120363, 9969601, mi13060851, RAY20221595}.\\
\hline
Reliability & TinyML allows IoT devices to operate reliably without constant reliance on cloud services~\cite{RAY20221595, DUTTA2021100461, 10177729}. \\
\hline
Enhanced Availability & TinyML prevents STI caused by cloud service interruptions or IoT layer crashes, which ensures continuous operation even during disruptions~\cite{DUTTA2021100461, kallimani2023tinyml}. \\
\hline
Information Filtering & TinyML enables filtering redundant information within devices, improving efficiency in scenarios such as surveillance systems with multiple cameras~\cite{DUTTA2021100461}. \\
\hline
\end{tabular}
\label{tab:tinyml-advantages}
\end{table*}

The expansion of the Internet has led to a significant increase in data generation. Data science addresses this challenge by employing statistical methods, data mining, and computational techniques to extract relevant information from large datasets. ML plays a vital role in this process by enabling computers to learn from training data and identify patterns in similar datasets~\cite{injadat2021machine}.

Despite their utility, traditional ML models often encounter limitations such as large size, computational complexity, and high energy consumption. These factors make them less suitable for deployment on resource-constrained edge devices. A common approach to mitigate this issue is offloading complex computational tasks from sensor nodes to edge gateways. For example, in fall detection systems proposed by Queralta et al.~\cite{8768883}, AI algorithms are applied at the edge gateways. However, this method may not always be efficient, as the energy required to transmit data between devices can exceed the energy used to run an ML model on the device itself~\cite{immonen2022tiny, banbury2020benchmarking}.

In response to these challenges, TinyML has emerged as a specialized field focused on optimizing ML models for deployment on edge devices with limited resources. TinyML aims to facilitate the effective deployment and execution of ML models on devices with minimal computational power, memory, and energy requirements~\cite{SABOVIC2023100736, 10177729}.

The goal of TinyML is to enable ML inference on ultra-low-power devices that consume within a few milliwatt or less~\cite{tinyml, RAY20221595, banbury2020benchmarking, s23042344}. This approach enables the implementation of ML capabilities directly on the edge, allowing for decision-making processes independent of cloud or remote computing infrastructures. The significance of TinyML lies in its ability to operate on widely available and cost-effective microcontrollers~\cite{immonen2022tiny}. By conducting inference close to the sensor and on the device, TinyML enhances real-time threat detection and improves privacy measures. This is achieved by minimizing the need for data transfer over potentially insecure long-distance connections, thereby reducing the risk of data interception or manipulation~\cite{RAY20221595, fi14120363, banbury2020benchmarking}.

\subsubsection{TinyML Advantages}

\begin{itemize}

\item \textbf{Bandwidth Efficiency:} Integrating ML models into compact, battery-less IoT devices enables local data processing and autonomous decision-making. This approach eliminates the need for transmitting collected data to the cloud, reducing reliance on cloud-based computing power and bandwidth requirements. The term battery-less refers to devices that operate without conventional batteries by harnessing energy from environmental or other external sources~\cite{SABOVIC2023100736}. Although the cloud possesses significant computing power, offloading the entire computational task to the cloud results in large bandwidth requirements~\cite{abbas2017mobile}. TinyML minimizes reliance on cloud services, which leads to decreased data transmission and potentially reduces bandwidth consumption~\cite{mi13060851}. Thus, bandwidth efficiency can be improved as an advantage of TinyML over conventional IoT systems~\cite{DUTTA2021100461}.

\item \textbf{Latency Reduction:} Minimizing dependence on external communication mitigates the problem of latency~\cite{abbas2017mobile, shaer2023corrfl}. TinyML significantly improves latency efficiency by enabling on-device analysis, minimizing delays from transferring data to cloud servers and ensuring swift decision-making, crucial for time-sensitive applications such as health care and autonomous vehicles~\cite{fi14120363, 10177729, RAY20221595, DUTTA2021100461}.

\item \textbf{Cost Savings:} TinyML contributes to cost savings in two distinct ways. First, by reducing data traffic and bandwidth requirements, TinyML minimizes the expenses associated with data transmission and storage~\cite{10177729, DUTTA2021100461}. This reduction in bandwidth usage directly translates into lower operational costs for IoT systems. Second, TinyML's implementation on affordable microcontrollers highlights its advantage in hardware costs. Unlike more resource-intensive IoT devices that rely heavily on costly cloud resources, TinyML operates efficiently on low-cost, resource-limited hardware~\cite{9166461, mi13060851}.

\item \textbf{Energy Efficiency:} Processing tasks consume significantly lower energy than wireless transmissions~\cite{9166461}. Consequently, executing computations directly on devices uses less energy than sending data wirelessly. This energy efficiency represents a key benefit of implementing TinyML on Microcontroller Units (MCUs)~\cite{9166461, RAY20221595}. Unlike powerful processors and Graphics Processing Units (GPUs) that require significant power consumption, IoT devices running on MCUs consume reduced energy~\cite{mi13060851, 10177729}. This allows for the deployment of IoT devices in various locations without the need for a constant power supply~\cite{mi13060851}.

\item \textbf{Enhanced Security and Privacy:} By minimizing data flow, security and privacy are enhanced, providing inherent protection by default~\cite{DUTTA2021100461}. Specifically, frequent cloud access raises privacy concerns and challenges the autonomy of edge devices~\cite{abbas2017mobile}. Minimized data transmission reduces the risk of malicious attacks; as a result, data security and privacy are naturally embedded into the core design of TinyML~\cite{fi14120363, 9969601, mi13060851, RAY20221595}.

\item \textbf{Reliability:} Reduced connectivity dependency enables IoT devices to operate reliably without constant reliance on cloud services while delivering accurate ML capabilities~\cite{RAY20221595}. This makes it an appealing choice for IoT applications that require cost-effective solutions, particularly in settings with limited connectivity ~\cite{DUTTA2021100461}. TinyML models can even operate without an Internet connection, unlike cloud ML models that rely on connectivity~\cite{10177729}.

\item \textbf{Enhanced Availability and Responsiveness:} TinyML significantly mitigates the issue of Short Time Inoperability (STI), commonly caused by cloud service interruptions or crashes in the IoT layer. Such incidents can lead to significant malfunctions in critical-risk environments. By enabling analytics directly on the IoT device, TinyML ensures continuous operation, even during external disruptions. As a result, TinyML contributes to improved availability and responsiveness in IoT systems~\cite{DUTTA2021100461, kallimani2023tinyml}.

\item \textbf{Information Filtering:} TinyML enables filtering redundant information, particularly in high data traffic situations. An example of this is in surveillance systems equipped with multiple cameras. By filtering out redundant images directly on the device, rather than transmitting all data to the cloud, TinyML enhances efficiency and ensures that only meaningful information is processed and sent for further analysis~\cite{DUTTA2021100461}.

\end{itemize}

The advantages of TinyML are summarized in Table~\ref{tab:tinyml-advantages}.

\begin{table*}
\centering
\caption{Applications of TinyML}
\renewcommand{\arraystretch}{1.25} 
\label{tab:tinyml-applications}
\begin{tabular}{|M{3.5cm}|L{13cm}|}
\hline
\multicolumn{1}{|c|}{\textbf{Application}} & \multicolumn{1}{c|}{\textbf{Description}}
\\
\hline
Wearable Technology & Enhancing health, safety, and communication through low-cost TinyML sensors for condition monitoring and beyond~\cite{immonen2022tiny}. \\
\hline
Healthcare Diagnostics & Improving hearing aid technology, aiding in health diagnosis, such as identifying COVID-19 symptoms~\cite{kallimani2023tinyml}. \\
\hline
Industrial Asset Monitoring & Leveraging TinyML for anomaly detection in extreme conditions, such as underwater pumps in wastewater management~\cite{s23042344}. \\
\hline
Plastic Component Analysis & Introducing a sensor system with TinyML for detecting artifacts and anomalies in plastic components~\cite{9969601}. \\
\hline
Autonomous Robotics & Utilizing TinyML in low-cost, resource-constrained autonomous robots for tasks such as search and rescue, and infrastructure monitoring~\cite{neuman2022tiny}. \\
\hline
Urban Air Quality Monitoring & Integrating TinyML into embedded systems for estimating CO2 emissions from vehicles, enhancing accuracy without cloud reliance~\cite{andrade2022tinyml}. \\
\hline
Indoor Positioning Systems & Improving location determination within indoor environments by classifying RSSI data with TinyML~\cite{s23031542}. \\
\hline
Wearable Deep Learning & Classifying human activities from wearable sensor data using on-device DL inference mechanisms~\cite{10.1145/3571306.3571415}. \\
\hline
Augmented Reality & Enabling real-time processing for AR glasses by avoiding cloud-based delays~\cite{fi14120363, banbury2020benchmarking}. \\
\hline
Ecological Conservation & Monitoring wildlife, such as sea turtles, for conservation purposes~\cite{kallimani2023tinyml}. \\
\hline
Autonomous Vehicles & Facilitating smarter navigation for autonomous vehicles~\cite{kallimani2023tinyml}. \\
\hline
Miscellaneous Applications & Including sign language detection, handwriting recognition, face mask identification, gesture interpretation, and speech transcription~\cite{mi13060851}. \\
\hline
\end{tabular}
\end{table*}

\subsubsection{TinyML Applications}

TinyML's real-time analytics and small computational footprint make it particularly suitable for a wide range of applications including IoT devices, wearable technology, autonomous robotics, and embedded systems. This transformative technology has the potential to reshape industries such as industrial monitoring, healthcare diagnostics, environmental conservation, and smart city development.

\begin{itemize}
    \item Wearable technology highlights TinyML's role in enhancing health, safety, and communication. Applications extend to condition monitoring in mobile machinery, where low-cost TinyML sensors offer solutions beyond expensive equipment~\cite{immonen2022tiny}.
    \item In healthcare, TinyML aids in improvements in hearing aid technology by distinguishing desired sounds in noisy environments. It also assists in health diagnosis, such as identifying COVID-19 symptoms~\cite{kallimani2023tinyml}.
    \item In industrial contexts, TinyML facilitates effective asset monitoring under extreme conditions. Antonini et al.~\cite{s23042344} demonstrated its use in detecting anomalies in underwater pumps at a wastewater management plant. The ESP32 microcontroller-based IoT Kit implemented a TinyML pipeline for data processing, from sampling to anomaly notification. Additionally, Albanese et al.~\cite{9969601} introduced a sensor system with three TinyML cameras on MCUs for detecting artifacts and anomalies in plastic components, showcasing its potential in industrial applications.
    \item The utility of TinyML in robotics is evident in the work of Neuman et al.~\cite{neuman2022tiny}, who explored deploying ML on low-cost, resource-constrained autonomous robots for tasks such as emergency search and rescue, and routine infrastructure monitoring.
    \item TinyML plays a critical role in urban air quality monitoring by integrating into embedded systems to estimate CO2 emissions from vehicles using soft-sensor techniques. This approach enhances accuracy without relying on cloud-based servers~\cite{andrade2022tinyml}.
    \item In indoor environments, where traditional methods such as GPS are inadequate, TinyML proves beneficial. Avellaneda et al.~\cite{s23031542} used TinyML to classify Radio Signal Strength Indicator (RSSI) data for improved accuracy in indoor positioning systems.
    \item BandX, a TinyML system, exemplifies the advantages of on-device DL inference mechanisms. Used in wearable sensors, BandX classifies human activities while ensuring greater network independence, enhanced data privacy, reduced power consumption, and lower latency and bandwidth needs~\cite{10.1145/3571306.3571415}.
    \item TinyML's impact extends to emerging technologies such as Augmented Reality (AR) glasses, where real-time demands make cloud or server calculations impractical. The technology’s ability to operate with always-on functionality and limited battery power is crucial for these applications~\cite{fi14120363, banbury2020benchmarking}.
    \item TinyML's applications in ecological conservation, autonomous vehicles, and Brain-Computer Interfaces highlight its versatility. It aids in tasks such as tumor detection, emotional intelligence, smart navigation, and monitoring wildlife, such as sea turtles~\cite{kallimani2023tinyml}.
    \item TinyML is also used for recognizing sign language and handwriting, identifying medical face masks, interpreting gestures, and transcribing speech~\cite{mi13060851}.
\end{itemize}


The summary of TinyML applications is presented in Table~\ref{tab:tinyml-applications}.
\begin{table*}
\centering
\caption{Summary of Cyber Threats and Attacks}
\renewcommand{\arraystretch}{1.25} 
\begin{tabular}{|M{1.75cm}|L{15.5cm}|}
\hline
\multicolumn{1}{|c|}{\textbf{Threat/Attack}} & \multicolumn{1}{c|}{\textbf{Description}}
\\
\hline
DoS and DDoS & DoS attacks aim to disable or restrict network services, while DDoS magnifies the attack by using multiple compromised devices to target a single system~\cite{DENEIRA2023109553}. \\
\hline
Port Scan & Attackers probe network hosts to gather details about port numbers, operating systems, and applications~\cite{ono2021proposal}. \\
\hline
Botnet & A network of malware-infected hosts controlled by malicious actors, often used for DDoS attacks and other malicious activities~\cite{fi13080198}. \\
\hline
Infiltration & Unauthorized actions on network hardware or software systems, detected by IDSs~\cite{s23135829}. \\
\hline
Web Attacks & Various web-based attacks, including Brute Force, XSS, and SQL Injection~\cite{park2021network, kaur2014study, 8854182}. \\
\hline
Heartbleed & Exploits a vulnerability in OpenSSL to steal sensitive data from servers~\cite{amodei2023measurement}. \\
\hline
Phishing & Deceives individuals into sharing sensitive data using fraudulent websites, emails, and social engineering tricks~\cite{hazell2023large}. \\
\hline
Ransomware & Malware that blocks access to personal data until a ransom is paid~\cite{BEAMAN2021102490}. \\
\hline
MitM attack & Attackers intercept and sometimes alter communication between two parties~\cite{7442758}. \\
\hline
Zero-Day & Attacks target vulnerabilities not yet known to the public or cybersecurity community~\cite{GUO2023175}. \\
\hline
\end{tabular}
\label{tab:cyber-threats}
\end{table*}

\subsection{Cybersecurity Threats}

Cybersecurity is crucial for protecting computer systems, networks, and data from digital attacks. This section provides an overview of key cyber threats that impact various domains, including Electric Vehicle Charging Infrastructures (EVCIs).

\subsubsection{DoS and DDoS (Distributed Denial-of-Service) Attacks} These attacks aim to disable or restrict network services by overwhelming the system with excessive traffic or exploiting system weaknesses, which makes the service inaccessible to legitimate users~\cite{DENEIRA2023109553}.
\subsubsection{Port Scan} Attackers probe network hosts to gather details about port numbers, operating systems, and applications by sending specific data and then analyzing the host's response~\cite{ono2021proposal}.
\subsubsection{Botnet} A botnet is a network of malware-infected hosts controlled by malicious actors. These hosts, known as bots, are typically managed by a Command and Control (C\&C) server, which can organize distributed attacks on other systems~\cite{fi13080198}.
\subsubsection{Infiltration} Infiltration refers to unauthorized actions on network hardware or software systems. The primary objective of Intrusion Detection System (IDS) is to detect and prevent these unauthorized actions~\cite{s23135829}.
\subsubsection{Web Attacks} Including Brute Force, Cross-Site Scripting (XSS), and SQL Injection, these attacks target vulnerabilities in web services to gain unauthorized access or compromise data integrity~\cite{park2021network, kaur2014study, 8854182}.

\subsubsection{Heartbleed} This attack undermines the encryption mechanism intended to secure online communications. Servers using a buggy version of Open Secure Socket Layer (OpenSSL) allow attackers to read the memory of systems, leading to the exposure of sensitive data such as private keys, user passwords, and confidential information~\cite{amodei2023measurement}.   
\subsubsection{Phishing and Spear Phishing} Phishing is a cyberattack strategy that deceives people into sharing sensitive data by using fraudulent websites, emails, and social engineering tricks. This leads victims to reveal personal, financial, or confidential information, often resulting in harmful consequences such as financial losses or data breaches~\cite{alkhalil2021phishing, hazell2023large}.
\subsubsection{Ransomware} Malware that carries out malicious actions such as blocking access to personal data until a ransom is paid. This ransom is often demanded in cryptocurrency such as Bitcoin, making it challenging to trace transactions and evading law enforcement~\cite{BEAMAN2021102490}.
\subsubsection{Man-in-the-Middle (MitM)} Attacks intercepting communications between two parties to steal or manipulate data~\cite{7442758}.
\subsubsection{Zero-Day Attacks} A novel cyber threat targeting vulnerabilities unknown to the public and the cybersecurity community. Traditional signature-based detection methods often fail against these attacks, as they depend on known attack signatures, which zero-day threats lack. However, ML can effectively detect these attacks by analyzing statistical patterns in data. This approach, not dependent on known attack signatures, is better suited for combating novel threats~\cite{GUO2023175}.


The Summary of cyber threats and attacks is presented in Table~\ref{tab:cyber-threats}. TinyML holds potential in addressing various challenges in cybersecurity and network vulnerabilities. It opens new possibilities for detecting, preventing, and mitigating threats. The subsequent section describes the capabilities and applications of TinyML in tackling these challenges.

\subsection{The Role of TinyML in Enhancing Cybersecurity}
As the IoT systems grow, the complexity of cyber threats increases, thus amplifying the importance of mitigating these threats. Unauthorized activities that threaten the confidentiality, integrity, and availability of these systems, known as intrusions, must be counteracted~\cite{khraisat2019survey}. For this purpose, IDSs are crucial for detecting unauthorized access, misuse, and breaches in computer networks and systems. They operate by monitoring and analyzing network traffic or system activities for malicious patterns or anomalies that indicate potential security threats~\cite{ozkan2021comprehensive}. There are two different types of IDSs: Signature-based Intrusion Detection System (SIDS) and Anomaly-based Intrusion Detection System (AIDS)~\cite{khraisat2019survey}.

SIDSs operate by matching patterns to identify known attacks, comparing intrusion signatures against a pre-existing database~\cite{khraisat2019survey}. Incorporating supervised ML algorithms in SIDS enhances their capability to recognize known threats. The key limitation here is that supervised learning models are trained on labeled datasets and face challenges with new, unfamiliar threats. Maintaining and processing a large database for SIDS may require considerable computational resources~\cite{abd2023iot}. Additionally, the centralized structure, common in traditional cybersecurity systems such as SIDS, may introduce certain vulnerabilities, including single points of failure and privacy issues~\cite{pourmirza2023cybersecurity}.

In contrast, AIDSs do not rely solely on known patterns. Instead, this approach uses ML and statistical methods to establish a baseline of normal behavior for a system. Deviations from this baseline are flagged as potential intrusions~\cite{khraisat2019survey}. ML models, especially unsupervised learning models, are capable of autonomously identifying patterns and anomalies, independent of predefined threat criteria. These models contribute to the creation of sophisticated frameworks for detecting malware, classifying spam, and identifying intrusions and anomalies. To facilitate this process, powerful servers with extensive computational resources are used to analyze large datasets and identify vulnerabilities~\cite{9969601}.

Furthermore, AIDSs, often referred to as behavior-based IDSs, exhibit a high False Alarm Rate (FAR). This problem can lead to unnecessary panic, wasted resources, and potential system downtime while verifying the authenticity of the alert. Additionally, the increasing prevalence of zero-day attacks, which exploit unknown vulnerabilities, poses a significant challenge to the effectiveness of AIDS. These attacks require the IDS to swiftly adapt, a task that is particularly challenging due to the inherent nature of these unforeseen threats~\cite{abd2023iot}.

Several papers explore various ML methods for IDS. For instance, Vibekananda et al. describe an IDS that employs a deep auto-encoder and multiple deep decoders for unsupervised classification. This system starts by processing raw, unlabelled network data, encompassing diverse network traffic features. The deep auto-encoder's task is to condense this data into a simpler form, retaining important characteristics. The deep decoders then work to rebuild the original data from this simplified version. In this reconstruction process, the system learns to identify anomalies or unusual patterns in network behavior that could indicate a threat~\cite{dutta2022unsupervised}. 

In another study, Bhoopesh et al. presents an IDS that utilizes XGBoost, an ML method known for its tree boosting technique. In this system, each decision tree is built sequentially, addressing errors identified by previous trees. The strength of this system lies in XGBoost's ability to effectively balance bias and variance, thus enabling precise predictions without the risk of overfitting~\cite{bhati2021improved}. Extending the application of ML in IDS, Mohamed et al. focus on employing several ML classifiers for IDS purposes within the context of Electric Vehicle Charging Stations (EVCSs)~\cite{elkashlan2023machine}.

\begin{table*}
\centering
\caption{Comparison between TinyML MCUs and Conventional ML Systems}
\renewcommand{\arraystretch}{1.25} 
\label{table1}
\begin{tabular}{|c|c|c|c|}
\hline
\textbf{Device} & \textbf{Memory Constraint} & \textbf{Device and CPU Specifications} & \textbf{Suitability} \\
\hline
STM32L432KC & 256 KB Flash, 64 KB SRAM & Arm® Cortex-M4, up to 80 MHz & TinyML \\ \hline
Arduino Nano 33 BLE & 1 MB Flash, 256 KB RAM & Arm® Cortex-M4F, 64 MHz & TinyML \\ \hline
ESP32 & 448 KB ROM, 520 KB SRAM & Xtensa Dual-Core, 240 MHz & TinyML \\ \hline
Raspberry Pi 4 & Up to 4 GB RAM, SD Card storage & Quad core ARM® Cortex-A72, 1.5GHz & Conventional ML \\ \hline
iPhone 12 Pro & 6GB RAM, Up to 512 GB storage & Hexa-core (2x3.1GHz Firestorm+4x1.8GHz Icestorm) & Conventional ML \\
\hline
\end{tabular}
\label{tab1}
\end{table*}

In the realm of EVCI, which is also the focus of this paper, the application of IDS is pivotal for maintaining cybersecurity and operational integrity. The work of ElKashlan et al.~\cite{elkashlan2023intrusion} highlights the effectiveness of ML classifiers in detecting DDoS attacks within the EVCS network; this research implements a variety of ML techniques to analyze an IoT dataset, aimed at identifying fraudulent traffic patterns specific to EVCSs. This approach improves the EVCS's defenses against cyber threats using IDS solutions.

Moving beyond the conventional ML approaches, TinyML stands out as a transformative technology in the field of IDS within resource-constrained environments. TinyML can enhance certain aspects of SIDS by making them more resource-efficient and enabling edge processing. Through the optimization of ML models, TinyML allows for the deployment of more advanced algorithms on edge devices. This could lead to the development of optimized anomaly detection systems that can detect new threats. Furthermore, TinyML facilitates on-device processing, reducing reliance on continuous communication with a central server. This enhancement not only conserves bandwidth but also accelerates response times, making the system more effective in real-world applications.

TinyML presents a versatile solution for enhancing both SIDS and AIDS in cybersecurity. By optimizing ML models for edge devices, TinyML enables the deployment of advanced algorithms that improve the efficiency and responsiveness of SIDS, facilitating resource-efficient edge processing and faster response times. Simultaneously, TinyML's adaptive and learning-based capabilities are ideal for AIDS, as they continuously monitor, analyze, and detect anomalies in real-time, directly on the devices. This approach reduces the reliance on network connectivity, minimizes latency, and boosts the overall resilience of cybersecurity systems. TinyML can adapt to evolving threats by employing advanced detection techniques, effectively countering new cybersecurity challenges.

\section{TinyML Challenges and Constraints}

TinyML brings significant advantages and potential in cybersecurity, but it faces its unique challenges. As the field grows, these challenges become evident, ranging from hardware limitations to complexities in both data and model management. Addressing these challenges is crucial for the effective and reliable deployment of TinyML systems. This section will thoroughly explore these challenges.

\subsection{Hardware Resource Constraints}

\subsubsection{Power Consumption}

While TinyML technology is praised for its low power consumption, making it ideal for smart devices aimed at operating autonomously in various locations, managing power consumption remains a critical challenge. Optimizing these systems for even greater efficiency and extending operational life without maintenance is a primary focus within the field~\cite{warden2018future}. Even the largest TinyML devices consume significantly less power than the smallest conventional ML devices~\cite{banbury2020benchmarking}. The literature widely recognizes that power consumption represents the primary constraint in the domain of TinyML~\cite{dutta2021implementation, RAY20221595, DELNEVO2023100729, 9014355, immonen2022tiny, computers12020023}.

MCUs such as the STM32L432KC, Arduino Nano 33, and ESP32 are examples of devices with minimal power requirements, operating at the milliwatt level. In contrast, instances of conventional ML systems, such as the Raspberry Pi 4 and the iPhone 12 Pro, demonstrate notably higher power consumption, on the order of watts.

\subsubsection{Memory Constraints}
While it is technically feasible to implement TinyML models on more resource-abundant hardware, such an approach could depart from the core principles of TinyML, which prioritizes minimal resource usage for maximum efficiency and sustainability. TinyML systems often face strict memory limitations, typically just a few kilobytes, due to their compact size~\cite{mi13060851}. In contrast, conventional ML systems such as smartphones handle resources in the range of a few gigabytes. However, TinyML systems generally operate with resources that are significantly smaller, often two orders of magnitude less~\cite{banbury2020benchmarking, fi14120363}. Current TinyML edge platforms are equipped with onboard flash memory of less than 1 megabytes~\cite{kallimani2023tinyml}.


Table~\ref{table1} outlines memory constraints for the STM32L432KC, Arduino Nano 33, and ESP32 MCUs. In contrast, the Raspberry Pi 4 and iPhone 12 Pro, representing conventional ML systems, exhibit comparatively greater memory capacities. This table emphasizes the significant disparity in memory resources between the two categories of devices.

\subsubsection{Computational Resources}
To adapt to the constraints of MCU-class devices, TinyML models must be both compact and efficient. This adaptation is necessary due to the limited embedded computing capacity and the processor's clock frequency, which typically falls in the MHz range~\cite{mi13060851, fi14120363}. These limitations affect both the size of the inputs that can be processed and the number of layers in the models~\cite{fi14120363}.

Table~\ref{table1} presents a comparison of the CPU specifications found in the previously mentioned TinyML MCUs and traditional ML systems. Each MCU model has its own unique set of CPU requirements, which define its fundamental characteristics. These requirements encompass the architecture and clock speed of the CPU. The table highlights the computational constraints faced by TinyML solutions due to their typically lower clock frequencies and less powerful CPU architectures.

\subsubsection{Software Limitation}

TinyML applications can be implemented on various platforms, including Linux/embedded Linux, and can even utilize cloud-enabled software solutions. However, this diversity implies that there is no universal solution~\cite{kallimani2023tinyml}. Software developers need to optimize applications for each specific platform which often leads to increased development time and resources.

As the field of TinyML evolves rapidly, there is an urgent need for a robust evaluation framework for software. While hardware benchmarks are important, a consistent software evaluation methodology is equally essential. Such a framework would allow for comparisons between different software solutions~\cite{fi14120363, DUTTA2021100461}.

\subsection{Data-Related Challenges}

\subsubsection{Data Privacy}

TinyML offers inherent security and data privacy advantages by keeping data within the device, thereby reducing data flow and exposure to attacks~\cite{TEKIN2023100670}. Numerous studies have emphasized the benefits of avoiding data transmission over shared mediums to enhance security and privacy~\cite{9014355, 9870017, s23031365, DELNEVO2023100729, 10.1145/3571306.3571415, mi13060851, 9166461}.

TinyML is not completely immune to over-the-air transmissions despite its focus on on-device data processing, particularly for tasks such as updates or selective data sharing. This becomes a concern when handling sensitive information such as health metrics~\cite{ahmed2022tinycare, s23031365} or location specifics~\cite{avellaneda2023tinyml}. Even these occasional transmissions can make security and privacy key challenges in the TinyML domain.

\subsubsection{Limited Datasets}

A key issue in TinyML is the scarcity of open-source datasets specifically designed for it~\cite{immonen2022tiny}. These datasets should be precise in time and space to match sensor-generated data and reflect the characteristics of low-power edge devices. Datasets that can handle data diversity and noise are crucial for effectively training TinyML systems~\cite{fi14120363}.

Banbury et al.~\cite{banbury2020benchmarking} reference several open-source datasets, such as Speech Commands for audio wake words~\cite{warden2018speech}, Visual Wake Words for visual cues~\cite{chowdhery2019visual}, CIFAR-10 for image classification~\cite{cifar10}, and ToyADMOS for anomaly detection~\cite{koizumi2019toyadmos}. Similarly, Alajlan et al.~\cite{mi13060851} list TinyML-utilized datasets, including ImageNET and VWW for images, physiological metrics datasets, road and traffic prediction datasets, and the Google Speech Commands dataset for keyword spotting. However, the common challenge is that these datasets are not specifically tailored for TinyML~\cite{banbury2020benchmarking}.

Additionally, there are concerns regarding the use of synthetic data~\cite{immonen2022tiny}. While synthetic data generated in controlled laboratory environments can help overcome dataset limitations, it presents unique challenges. Models trained on such synthetic data may face difficulties when applied to real-world scenarios. As a result, unforeseen issues and practical difficulties can arise~\cite{DELNEVO2023100729}.

\begin{table*}
\centering
\caption{Summary of TinyML Challenges}
\renewcommand{\arraystretch}{1.25} 
\label{table-tinyml-challenges}
\begin{tabular}{|M{4cm}|L{0.7\linewidth}|}
\hline
\multicolumn{1}{|c|}{\textbf{Category}} & \multicolumn{1}{c|}{\textbf{Challenges}}
\\
\hline
Hardware Resource Constraints & 
\begin{itemize}
  \vspace{5pt}
  \item \textbf{Power Consumption}: TinyML aims for ultra-low-power consumption operating at the milliwatt level, significantly lower than conventional ML systems which operate at the watt scale~\cite{banbury2020benchmarking, s23042344, dutta2021implementation, RAY20221595, DELNEVO2023100729, 9014355, immonen2022tiny, computers12020023}.
  \vspace{5pt}
  \item \textbf{Memory Constraints}: TinyML systems operate with a few KBs of memory, whereas conventional ML systems have resources on the order of a few GBs~\cite{mi13060851, banbury2020benchmarking, fi14120363}.
  \vspace{5pt}
  \item \textbf{Computational Resources}: TinyML requires compact and efficient models due to limited processor clock frequency and embedded computing capability~\cite{mi13060851, fi14120363}.
  \vspace{5pt}
  \item \textbf{Software Limitation}: Lack of a one-size-fits-all solution which necessitates platform-specific optimizations and a robust evaluation framework for software~\cite{kallimani2023tinyml, fi14120363, DUTTA2021100461}.
\end{itemize} \\
\hline
Data-Related Challenges & 
\begin{itemize}
  \vspace{5pt}
  \item \textbf{Data Privacy}: Emphasizes on-device data processing to enhance security and privacy, although it is not entirely immune to over-the-air transmissions~\cite{9014355, 9870017, s23031365, DELNEVO2023100729, 10.1145/3571306.3571415, mi13060851, 9166461}.
  \vspace{5pt}
  \item \textbf{Limited Datasets}: Lack of open-source datasets tailored for TinyML, with existing datasets not exclusively designed for TinyML applications~\cite{immonen2022tiny, fi14120363}.
\end{itemize} \\
\hline
Model-Related Challenges & 
\begin{itemize}
  \vspace{5pt}
  \item \textbf{Lack of Unified Framework}: The absence of a generic, vendor-neutral TinyML framework impedes the standardization and performance assessment across diverse hardware and software~\cite{immonen2022tiny, mi13060851, david2021tensorflow}.
  \vspace{5pt}
  \item \textbf{Robustness}: Refers to a model's capacity to sustain performance despite perturbations or noise, an area that remains largely unexplored in TinyML~\cite{sun2022improving}.
  \vspace{5pt}
  \item \textbf{AutoML Deployability}: Existing AutoML frameworks are limited in their deployability on resource-constrained devices~\cite{PEREGO2022117876}.
  \vspace{5pt}
  \item \textbf{Real-time Inflexibility}: TinyML is primarily designed for batch/offline settings, which creates challenges for real-time adaptability and widespread use in IoT applications~\cite{ren2021tinyol}.
\end{itemize} \\
\hline
\end{tabular}
\end{table*}

\subsection{Model-Related Challenges}

\subsubsection{Lack of Unified Framework}

The development of TinyML faces challenges due to the absence of a unified framework that can handle the extensive variety in hardware and software~\cite{immonen2022tiny, mi13060851}. With a broad range of devices available, each featuring distinct characteristics such as power usage, memory, and communication protocols, standardizing TinyML tools and benchmarks is a complicated task~\cite{9166461}.

For TinyML to truly thrive, a comprehensive framework is crucial. This framework should support model training on advanced platforms and deployment on embedded targets, as well as facilitate model orchestration, testing, and integration into device production~\cite{fi14120363}. However, the current landscape is characterized by a fragmented approach. Engineers often create custom frameworks for specific hardware platforms when deploying neural networks. This approach results in narrowly focused solutions that lack versatility for various applications and hardware compatibility~\cite{david2021tensorflow}.

\subsubsection{Robustness}
Robustness refers to a model's capability to maintain its performance in the presence of perturbations or noise. For instance, in the context of machine translation tasks, robustness is the ability of the model to adapt and perform well when faced with new and different language data~\cite{sun2022improving}.

Although the robustness of TinyML algorithms has not been extensively studied by researchers so far, it is a potential area for future research. Improving the robustness of TinyML models can help overcome the challenges of deploying ML models on resource-constrained devices that may encounter diverse environmental conditions or input data. Stronger robustness can boost the accuracy and dependability of TinyML in real-world uses, such as in medical devices or smart sensors.

\subsubsection{AutoML Deployability}
Automated Machine Learning (AutoML) is vital in ML as it automates data analytics tasks, thereby reducing reliance on domain expertise and human effort. By automating algorithm selection and hyperparameter tuning, AutoML enables efficient decision-making and saves valuable resources such as time, finances, and human effort~\cite{YANG2022105366}. Effective hyperparameter optimization, which is one of the central tasks of AutoML, offers a balance between complexity and generalization~\cite{yang2020hyperparameter}. While existing AutoML frameworks can generate highly accurate models, their deployability on resource-constrained devices often falls short.

In pervasive computing, ML models are deployed on small devices such as smart-home fixtures (light switches, door handles, windows) and personal devices (smartphones, smartwatches). These models are responsible for tasks including object and motion activity recognition. AutoML and Neural Architecture Search (NAS), while standard methods for model optimization, are computationally intensive for microcontroller devices. Therefore, optimizing AutoML to accommodate limited hardware capabilities is essential for effective deployment on these devices~\cite{PEREGO2022117876}.

\subsubsection{Real-time Inflexibility}

TinyML solutions are specifically designed to work in a batch/offline setting, where data is processed in batches rather than in real-time. The model is trained first on a powerful machine using a substantial amount of pre-collected data and then flashed to MCUs for inference process. As a result, the model becomes static and challenging to adapt to new data. This lack of flexibility is a significant obstacle to the extensive utilization of TinyML in IoT applications~\cite{ren2021tinyol}.

The summary of TinyML challenges can be found in Table~\ref{table-tinyml-challenges}. Addressing these challenges is crucial for the effective deployment of TinyML in EVCI. For example, EVs aim to minimize power consumption to extend battery life and enhance sustainable development. By optimizing TinyML solutions to operate within the stringent resource constraints of EVCI, we can enhance the security, efficiency, and reliability of these critical systems, ensuring they can meet the demands of future smart cities and the increasing adoption of EVs.

\section{Potential Solutions, Tools, and Libraries for TinyML}

In this section, we shift our focus to exploring viable solutions that aim to mitigate the various challenges previously identified in the realm of TinyML. These strategies present promising pathways for enhancing both the efficiency and reliability of TinyML implementations.

\subsection{Energy Harvesting}
Energy harvesting technology transforms environmental mechanical vibrations into electrical power~\cite{DIVYA2023108084}. Compact software is essential to ensure small power consumption and enable the implementation of TinyML. TinyML systems must operate within strict constraints while maintaining high accuracy. Energy harvesting techniques can be a solution for TinyML by providing power directly to its operations in edge devices or supporting battery-operated embedded systems~\cite{kallimani2023tinyml}. By integrating energy-harvesting mechanisms with TinyML systems, we are stepping closer to an era where devices can self-sustain and bridge the gap between computational demands and power constraints.

\subsection{Privacy-Preserving Solution}
Data privacy is a major concern for individuals, especially regarding the transmission and storage of their data. Kavya and Eric~\cite{kopparapu2021tinyfedtl} propose a decentralized learning mechanism, TinyFedTL, to address TinyML's security and privacy challenges. Using this method ensures that data stays localized on the device, thereby preventing centralization of personal data and enhancing security. Federated transfer learning offers a solution to the TinyML community that respects privacy constraints, minimizes communication overhead, and adapts to the hardware limitations of microcontrollers such as the Arduino Nano 33 BLE Sense.

\subsection{Computation Optimization}
Deploying complex models on microcontrollers is challenging due to their limited computational and energy capacities. As a solution, researchers are exploring lightweight frameworks, such as TensorFlow Lite (TFLite)~\cite{tensorflowlite}, which optimize model deployment on various edge devices. The process of converting a TensorFlow model to TFLite format involves several potential optimizations aimed at reducing the model's computational complexity and storage requirements. 

One of the key techniques is quantization, which lowers the precision of the numerical values representing the model's parameters, usually from 32-bit floating-point numbers to lower-bit integers. This technique results in a smaller model size and faster inference times but may have a slight impact on model accuracy. Another technique is pruning, which eliminates less important parameters within the neural network, leaving the essential ones that have a more significant impact on predictions. Clustering is another optimization technique that groups the weights of each layer in a model into a predefined number of clusters, sharing the centroid values for the weights belonging to each individual cluster. This technique reduces the complexity of the model by reducing the number of unique weight values which offers deployment benefits similar to pruning. There is also a possibility of operation fusion, where composite operations in TensorFlow can be converted to a single, fused operation in TFLite. This conversion to a fused operation streamlines the computational graph which reduces the number of primitive operations that need to be executed. Each optimization comes with its own trade-offs, often in the form of reduced accuracy or specific hardware requirements. 
However, these trade-offs are critical for achieving fast and efficient model inference on resource-constrained devices~\cite{tensorflow2021lite1, immonen2022tiny}.


Deploying models on embedded hardware presents unique challenges, including the need for portability, flexibility, and resource utilization efficiency. These systems often have strict constraints on memory, processing power, and energy consumption which makes standard deployment methods unsuitable. To meet these needs, David et al.~\cite{david2021tensorflow} present TensorFlow Lite Micro (TFLM), designed specifically for embedded hardware deployment. TFLM prioritizes portability and flexibility, offering a comprehensive deployment framework that addresses the limitations of resource-constrained devices. The process involves creating a neural network model, generating an operator resolver via the client API, providing memory, establishing an interpreter, and executing the model.

\subsection{Real-Time Flexibility}

Real-time flexibility is essential due to unpredictable variations in conditions, inputs, and requirements. For instance, in wearable devices and IoT systems, the context in which TinyML models operate can change as users interact with different objects or move around. Real-time adaptability enables these models to effectively handle new scenarios, unseen data distributions, and unexpected events, thereby ensuring resilient and dependable operation.

Through an interesting paper, Pavan et al.~\cite{pavan2023tybox} introduce TyBox, a toolbox designed to solve the challenges of incremental on-device learning in TinyML. Unlike traditional TinyML models where training is typically offloaded to the cloud, TyBox enables both the training and inference to occur directly on low-resource devices, such as embedded systems or IoT units. This capability is crucial for real-time adaptability, as TyBox's framework allows TinyML-enabled devices to learn and adapt while they are deployed in the field. This on-the-fly learning is vital for environments that are evolving or under different working conditions. For instance, TyBox could help a TinyML application adapt to a new gesture-recognition command, fine-tune vocal commands for specific users, or adapt to concept drifts, such as a change in lighting conditions for image recognition tasks.

\subsection{Model Generalization}

Transfer Learning (TL) in Deep Neural Networks (DNNs) involves transferring weights from a pre-trained model to a new dataset~\cite{yang2022transfer}. By enabling domain generalization, TL allows models to adapt to new scenarios without being retrained from scratch, and therefore saving valuable computational power. 

Supriya et al.~\cite{ASUTKAR2023119016} focus on enhancing the potential of TinyML for fault diagnosis in machinery through the use of Convolutional Neural Networks (CNNs) and TL. They found that by fine-tuning the parameters in the convolutional layers and transferring parameters in the dense layers, they could achieve significantly better domain generalization. This meant that the models were capable of adapting to new, unseen domains, which is crucial when the operating conditions or machine instances change.


\subsection{AutoTinyML}
As mentioned before, AutoML is a field that aims to automatically select, construct, tune, and update ML models for optimal performance on specific tasks~\cite{YANG2022105366}. To address the challenges of AutoML deployability on resource-constrained devices, Riccardo et al.~\cite{PEREGO2022117876} propose the innovative AutoTinyML framework. This framework enhances AutoML and Neural Architecture Search frameworks by integrating hardware constraints of tiny devices, such as microcontrollers, into the optimization process. The approach employs a two-phased Sequential Model-Based Optimization (SMBO) strategy. This strategy first estimates the feasible search space in terms of deployability and then optimizes within that space. This process enables the creation of precise ML models while taking into account the deployability constraints.


\subsection{Tools and Libraries}

Various TinyML tools and libraries have emerged to facilitate ML on edge devices~\cite{fi14120363}. These can be classified into different categories:

\subsubsection{Model Optimization and Conversion Tools}
\begin{itemize}
    \item TFLite~\cite{tensorflow_lite_models} offers a streamlined version of TensorFlow models for a variety of ML tasks with optimized size and efficiency.
    \item TFLM~\cite{tensorflow_lite_micro}, a derivative of TFLite, is engineered for extremely resource-constrained devices such as microcontrollers.
    \item STM32Cube.AI~\cite{STM32CubeAI} optimizes and deploys trained neural network models across STM32 microcontrollers. It also offers a graphical interface and a command line tool.
    \item ONNX Runtime~\cite{onnxruntime} accelerates ML inference and training across various frameworks and hardware, widely used in Microsoft products.
\end{itemize}

\subsubsection{Development Environments and Frameworks}
\begin{itemize}
    \item PlatformIO~\cite{platformio} is a comprehensive development environment that supports a wide range of microcontrollers and frameworks. It features an Integrated Development Environment (IDE) along with advanced debugging tools.
    \item NanoEdge AI Studio~\cite{NanoEdgeAIStudio} is a PC-based development environment which is more focused on ease of use for creating ML libraries for microcontrollers.
    \item PyTorch Mobile~\cite{pytorchmobile} provides an end-to-end workflow for deploying ML models on mobile devices.
    \item ELL~\cite{embeddedlearninglibrary}, by Microsoft Research, enables deployment of machine-learned models onto platforms such as Raspberry Pi and Arduino.
\end{itemize}
    
\subsubsection{Specialized Frameworks and Libraries}
\begin{itemize}
    \item uTensor~\cite{uTensor} is an embedded ML framework optimized for devices with limited memory.
    \item CMSIS NN Software Library~\cite{cmsisnn} offers neural network kernels optimized for Cortex-M processor cores.
    \item OpenMV~\cite{openmv}, often compared to the Arduino of Machine Vision, simplifies the implementation of machine vision algorithms.
\end{itemize}
    
\subsubsection{AI Integration and Industry-Specific Tools}
\begin{itemize}
    \item Edge Impulse~\cite{edgeimpulse} offers a cloud-based service for model development, particularly for sensor data collection and algorithm development for edge deployment.
    \item AI-CUBE~\cite{ai-cube} aims to optimize AI and Big Data use in the European process industry by providing a roadmap for various sectors.
\end{itemize}

In conclusion, a wide range of TinyML tools and libraries are available for ML on edge devices. These tools make it easier to run ML models on devices with limited resources, from mobile phones to industrial equipment.

\section{Case Study: Electric Vehicle Charging Infrastructure}

Among the numerous applications of TinyML across various sectors, EVCIs stand out as a particularly crucial and challenging domain. As the world moves towards smarter and more connected urban environments, the role of EVCI becomes increasingly significant in ensuring the seamless integration of electric vehicles into the urban landscape.

The drive for efficient, reliable, green, and connected smart cities has encouraged the use of EVs as the main future means of transportation~\cite{8994200, su11143863}. The anticipated increase in EVs creates a crucial need to expand the EVCS infrastructure~\cite{en13010176}. This expansion, aiming to meet the projected growth of EVs by 2030, requires adding 22 million charging points each year, which is 1.3 times more than the total number of charging points deployed to date~\cite{bibra2022global}. Therefore, EVCSs, as integral components of the IoT ecosystem in smart cities, are pivotal in supporting the widespread adoption of EVs.

However, the expansion of EVCSs introduces significant cybersecurity challenges. These systems are vulnerable to various threats, including tampering with charging processes through WiFi connections, bypassing authentication mechanisms, injecting malicious software, and disabling chargers~\cite{9272723}. Given these potential cyber threats, there is an urgent need for robust security solutions.

While traditional ML models provide a baseline defense against cyber threats in EVCSs, they often face resource constraints when deployed on IoT devices, which is a critical consideration in the expansion of EVCI~\cite{elkashlan2023machine}. In this context, TinyML emerges as a more suitable alternative. TinyML offers substantial advantages in terms of cost and energy effectiveness, while still providing enhanced security and privacy, as well as real-time processing capabilities, making it an optimal choice for improving the safety and efficiency of EVCIs.

\subsection{Framework and Standards}

The EV charging ecosystem represents a dynamic cyber-physical system that comprises interconnected hardware and software elements~\cite{sarieddine2022investigating}. In the context of our research, we specifically emphasize the software aspect of this ecosystem.

\begin{figure}[!t]
\centerline{\includegraphics[width=\columnwidth]{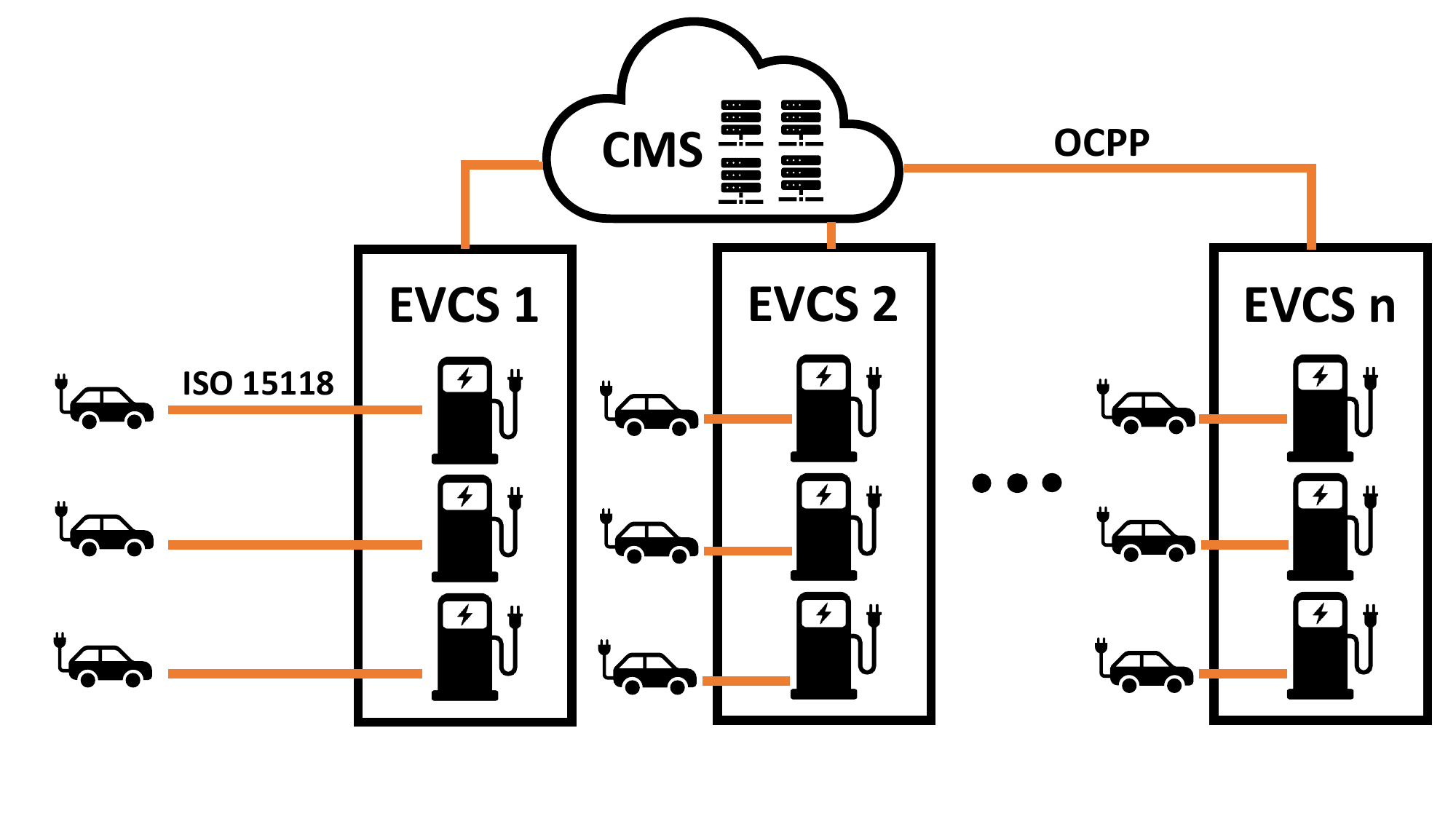}}
\caption{Communication scheme between EV and EVCS (ISO 15118) and between EVCS and CMS (OCPP)}
\label{fig:fig2}
\end{figure}

This section presents a framework that outlines how reliable communication channels are crucial for the smooth functioning of EVs. This framework consists of three main components: EV, EVCS, and the Central Management System (CMS). An EV relies on these systems to efficiently manage its charging process. An EVCS, comprising one or multiple Electric Vehicle Supply Equipment (EVSE) units, delivers power to EVs. The CMS serves as a cloud-based server with the primary role of monitoring and managing various charging stations. Its tasks involve scheduling and supervising the charging process, maintaining logs, managing both authorized and unauthorized transactions, and conducting remote diagnostics and adjustments~\cite{10.1145/3437258}. These components rely on robust communication channels to ensure efficient energy management, secure payment transactions, and effective coordination. Communication takes place between the EV battery management system and the EVCS, between the EVCS and the CMS, and among the CMS, energy suppliers, and the power grid~\cite{alcaraz2017ocpp}. Figure~\ref{fig:fig2} provides a schematic representation of this framework.

Various standards are utilized globally for EVCIs~\cite{ayob2014review, das2020electric}. ISO 15118, OCPP, SAE, CHAdeMO, and IEC are among the most well-known standards in the EV charging field. Here is a brief overview of each of these standards:

\subsubsection{ISO 15118} International Organization for Standardization (ISO) 15118 protocol is specifically designed for communication between EVs and EVCS~\cite{sarieddine2022investigating, WELLISCH201555, balakrishnan2023design} allowing them to negotiate and establish a proper charging schedule and protect against unauthorized access or tampering. ISO 15118 ensures secure communication between EVs and EVCS by preventing malicious third parties from intercepting and modifying messages or tampering with billing information. This is achieved through the introduction of the Plug\&Charge feature~\cite{metere2021securing}, which requires the establishment of a secure communication link between the EV and EVCS. This feature utilizes digital certificates for secure communication and enables automated authentication and authorization~\cite{metere2021securing}. This protocol also defines the structures and formats of the messages exchanged between EVs and EVCS. It specifies how EVs and EVCS should encrypt and decrypt messages which ensures the confidentiality and integrity of the communication~\cite{8994200}. ISO 15118 is part of the Combined Charging System (CCS) and promotes the adoption of mature security protocols such as Transport Layer Security (TLS). ISO 15118 establishes requirements for both the physical and datalink layers. This standard also includes a comprehensive security framework for enhancing the layers' security protocols~\cite{metere2021securing}.

\subsubsection{OCPP} Open Charge Point Protocol (OCPP) is an open and widely adopted communication protocol for EV charging stations~\cite{8994200}. OCPP facilitates the communication between a EVCS and a CMS~\cite{balakrishnan2023design, alcaraz2017ocpp}. For Communication between the EVCS and the CMS, a simple Hypertext Transfer Protocol Secure (HTTP/HTTPS) protocol is employed.  Given the wide array of EVCS operators and the critical importance of this communication, various operators have devised their own protocols. However, the OCPP is leading the efforts towards a standardized communication protocol in this domain~\cite{10.1145/3437258}. The primary functions of the OCPP protocol revolve around managing charging processes and reservations while considering security restrictions. Its main focus is to ensure that charging occurs only when authorized by a billing system, thus addressing potential fraudulent activities related to payment systems~\cite{alcaraz2017ocpp}. Sarieddine et al.~\cite{sarieddine2022investigating} also mention that the OCPP protocol defines two primary roles: a lightweight client representing the EVCS and a central server representing the CMS, and this protocol employs transaction functional blocks, where each entity requires a response for initiated transactions. The OCPP has evolved to prioritize security, especially in version 2.0~\cite{metere2021securing}. The protocol includes features such as secure firmware updates, security logging, event notification, and authentication. OCPP utilizes TLS for secure communication and implements key management for client-side certificates~\cite{metere2021securing}. Another key features of OCPP include support for EV-grid standards, remote control capabilities, and smart charging scenarios through charging profiles. With OCPP, the CMS can remotely start/stop charging processes, monitor and modify charging station status, and implement intelligent charging profiles, specifying power limits and duration for optimal charging management throughout the day~\cite{10.1145/3437258}.

\subsubsection{SAE} Society of Automotive Engineers (SAE) standards encompass both physical and communication protocols between the EV and the EVCS. The SAEJ2293 standard addresses power requirements, system architecture, and communication requirements for EV charging. SAEJ1772 specifies equipment ratings for Alternating Current (AC) and Direct Current (DC) charging, with different levels of charging available. SAEJ1773 focuses on inductively coupled charging schemes, including requirements for manual connection and software interface. SAEJ2847~\cite{balakrishnan2023design, falk2012electric} and SAEJ2836 define communication requirements and use cases between EVs and charging infrastructure. SAEJ2931 establishes digital communication requirements between EVs, EVSE, utility, and energy service interfaces. SAEJ2954 and the Recommended Practice (RP) version provide specifications for wireless charging and include features such as driving assistance and autonomous charging. These SAE standards ensure interoperability, safety, and efficient communication in EV charging systems~\cite{das2020electric}.

\subsubsection{CHAdeMO} CHAdeMO standard encompasses both physical and communication aspects. For the physical aspect, CHAdeMO defines the physical connector and charging specifications for high-voltage DC fast charging. The CHAdeMO protocol is used for communication between the EVCS and the EV, ensuring optimal and fast charging based on the EV's commands. CHAdeMO chargers are popular in Japan~\cite{10.1145/3437258} and Europe~\cite{ayob2014review}. CHAdeMO, while enabling Vehicle-to-Grid (V2G) operations, has certain limitations that raise concerns. From the communication side, one of the main drawbacks is the lack of secure communication features, relying instead on Controller Area Network (CAN) communication. CAN enables electronic components within a vehicle to communicate with each other without relying on a central computer. This exposes the vehicle's CAN bus to potential security risks, as unencrypted communication leaves it vulnerable to control or programming by malicious entities. Although CHAdeMO follows IEC standards for charging and digital communication, it currently lacks robust security measures. There is recognition of the importance of security, and efforts are being made to develop a unified communication protocol that would utilize Ethernet, Transmission Control Protocol/Internet Protocol (TCP/IP), PKI infrastructure, and TLS encryption for enhanced security, similar to how TLS is employed on the Internet ~\cite{metere2021securing}.

\subsubsection{IEC} The International Electrotechnical Commission (IEC) standards, such as IEC 61851, IEC 61980, and IEC 62196, covers physical and non-physical aspects of EV charging systems. These standards govern the protocols and interfaces used for communication between EVs and EVCS~\cite{10.1145/3437258}. These standards cover the overall operation, onboard and off-board equipment, and specific components used in EV charging. For example, IEC 61851 sets the standard for conductive charging systems, IEC 61980 addresses Wireless Power Transfer (WPT) systems, and IEC 62196 defines the specifications for plugs, connectors, and inlets used in conductive charging. These standards ensure the safety, compatibility, and reliability of the physical components involved in EV charging, providing guidelines for manufacturers, installers, and users~\cite{das2020electric}. However, a drawback of the IEC standard is that it does not provide specific standards for reverse power flow in EVs -reverse power flow refers to the capability of EVs to supply power back to the grid~\cite{10.1145/3437258}. 

There are also specific standards designed with a focus on safety in the context of EVs. Organizations such as National Fire Protection Association (NFPA) and National Electrical Code (NEC) have developed safety standards that are crucial for ensuring secure EV charging and grid integration. NFPA's standard NFPA 70 provides guidance on electrical equipment wiring and safety for EV charging which covers aspects such as electrical conductors and equipment within buildings, as well as connections to electricity supplies. These safety standards are instrumental in establishing a safe and reliable environment for EV charging infrastructure and minimizing potential hazards~\cite{das2020electric}.

While SAE, IEC, ISO, and ChAdeMO offer communication options between the EV and the EVCS/Grid, and OCPP addresses the communication between the EVCS and the CMS, in our specific scenario, ISO and OCPP are chosen and utilized. By choosing ISO and OCPP, there is a greater potential for harmonization and compatibility across different regions. Additionally, the use of standardized protocols reduces vulnerabilities caused by the diverse set of protocols and simplifies system management, minimizing the risk of exploitation and errors during integration~\cite{10.1145/3437258}.

\begin{figure}[!t]
\centerline{\includegraphics[width=\columnwidth]{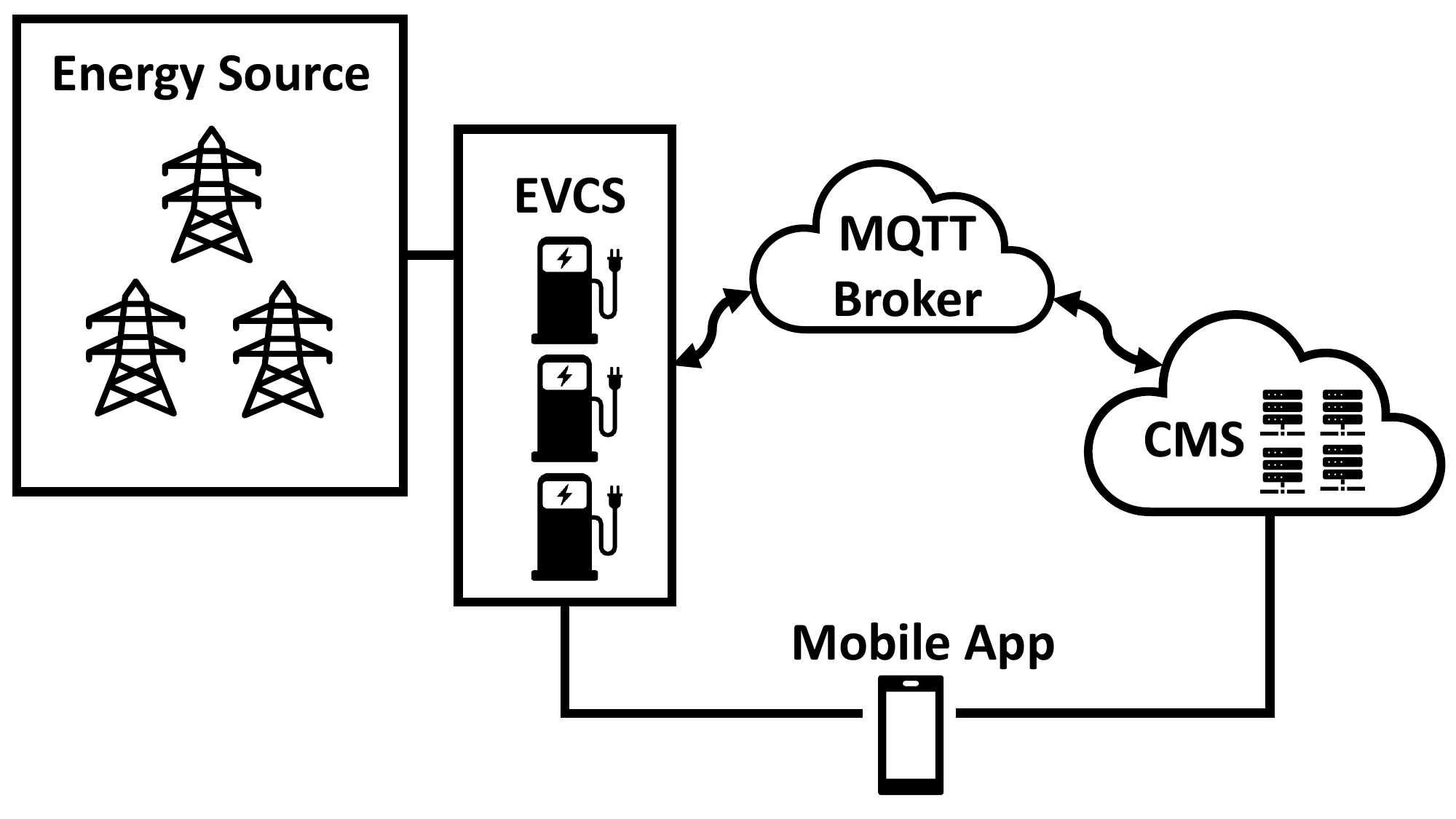}}
\caption{IoT-based smart charging station ~\cite{balakrishnan2023design}}
\label{fig:fig3}
\end{figure}

\subsection{Communication Frameworks for EV Charging}

In this subsection, various frameworks from different research papers are reviewed, along with the communication protocols employed in them. This analysis is conducted to gain a thorough understanding of the functionalities and interactions of these frameworks through their respective communication strategies.

\subsubsection{IoT-Enabled Smart Charging Network}
The smart EV charging infrastructure proposed by Balakrishnan et al. consists of several key components, namely the smart EVSE, Message Queuing Telemetry Transport (MQTT) broker, CMS, and a mobile application. The smart EVSE device is designed to intelligently charge EVs, monitor charging activity, and communicate with the CMS using MQTT protocol. The device utilizes a unique QR code for identification, controls the charging process, and ensures safety mechanisms are in place. The ESP8266 microcontroller is used to coordinate charging and communication activities, while a Wi-Fi module enables connectivity to the internet. The MQTT broker acts as a communication hub, facilitating the exchange of status information between the smart EVSE devices and the CMS. The CMS is responsible for managing and coordinating charging activities, recording energy usage, and providing additional services. The mobile application enables users to remotely start and stop charging sessions, monitor usage statistics, and access relevant information. This integrated architecture offers a contactless and efficient charging experience for EV owners~\cite{balakrishnan2023design}. Figure~\ref{fig:fig3} depicts the overall design and components involved. 

In this smart charging system, various messages are exchanged between the components to facilitate seamless charging operations. The smart EVSE device communicates with the web-based central management server using the MQTT protocol. The primary messages exchanged between the smart EVSE and the server include the \texttt{Boot\_Notification()} message, which is sent by the device upon power-up to notify the server of its presence and readiness. The \texttt{Device\_Healthy()} message is continuously published by the smart EVSE to indicate its operational status. The \texttt{Device\_Start()} message is used to initiate the charging session, while the \texttt{Device\_StopCharging()} message is sent to signal the completion of the charging process. These messages enable device registration, health monitoring, and control over the charging process. The CMS communicates with the smart EVSE through the MQTT broker, and messages such as \texttt{Remote\_start()} and \texttt{Remote\_stop()} are used for remote control of the charging sessions, allowing the server to start or stop charging on the device. Additionally, a mobile application is employed to initiate and monitor charging sessions, utilizing messages such as \texttt{Remote\_start()} to request the start of a charging session and \texttt{Remote\_stop()} to request the end of a charging session. The mobile application communicates with the management server, which then interacts with the smart EVSE using the MQTT broker. The mobile application also provides access to usage statistics and allows for a contactless charging experience~\cite{balakrishnan2023design}.

\begin{figure}[!t]
\centerline{\includegraphics[width=\columnwidth]{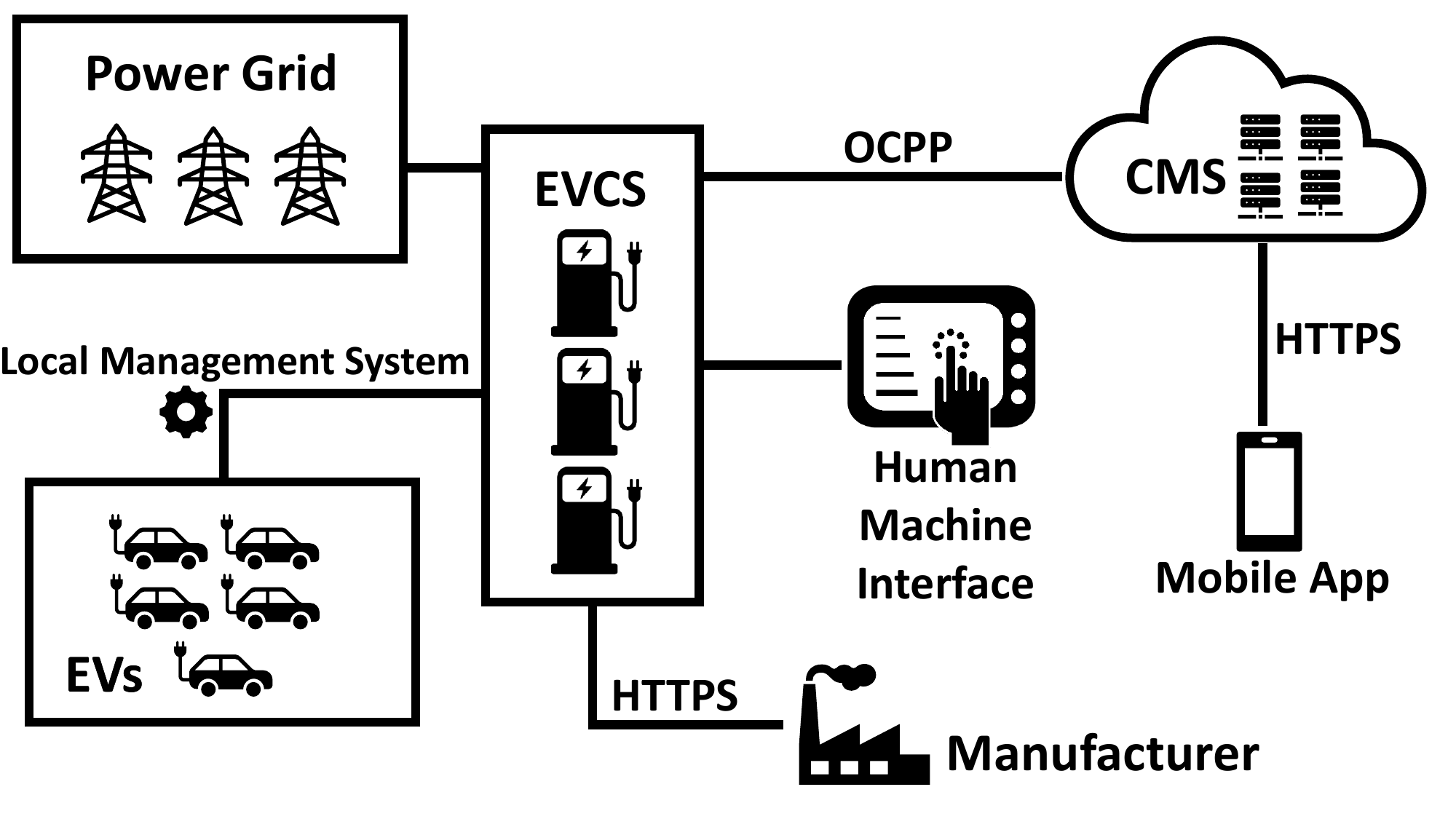}}
\caption{An overview of the interactions within the EV charging ecosystem ~\cite{sarieddine2022investigating}}
\label{fig:fig4}
\end{figure}

\subsubsection{Mobile Applications within the Charging Ecosystem}
Sarieddine et al.~\cite{sarieddine2022investigating} emphasize on the vulnerabilities present in EV charging mobile applications. These applications enhance user experience and system flexibility. Communication between the mobile applications and the CMS occurs over HTTPS, incorporating SSL/TLS for secure and encrypted communication. The mobile applications offer functionalities such as discovering nearby EVCS, remote start/stop of charging sessions, and scheduling charging. Popular mobile applications such as ChargePoint and ChargeHub provide remote monitoring and control of EVCS. 

Besides, The CMS provides API endpoints for communication between the mobile application and EVCSs. Each operator has their own CMS responsible for tasks such as reservation, scheduling, payments, and monitoring. The CMS communicates with the EVCS using the OCPP. Figure~\ref{fig:fig4} provides an overview of the interactions within the EV charging ecosystem proposed by Sarieddine et al.~\cite{sarieddine2022investigating}.

\subsubsection{Power and Communication Flows in Electric Vehicle Charging Infrastructure}

\begin{figure}[!t]
\centerline{\includegraphics[width=\columnwidth]{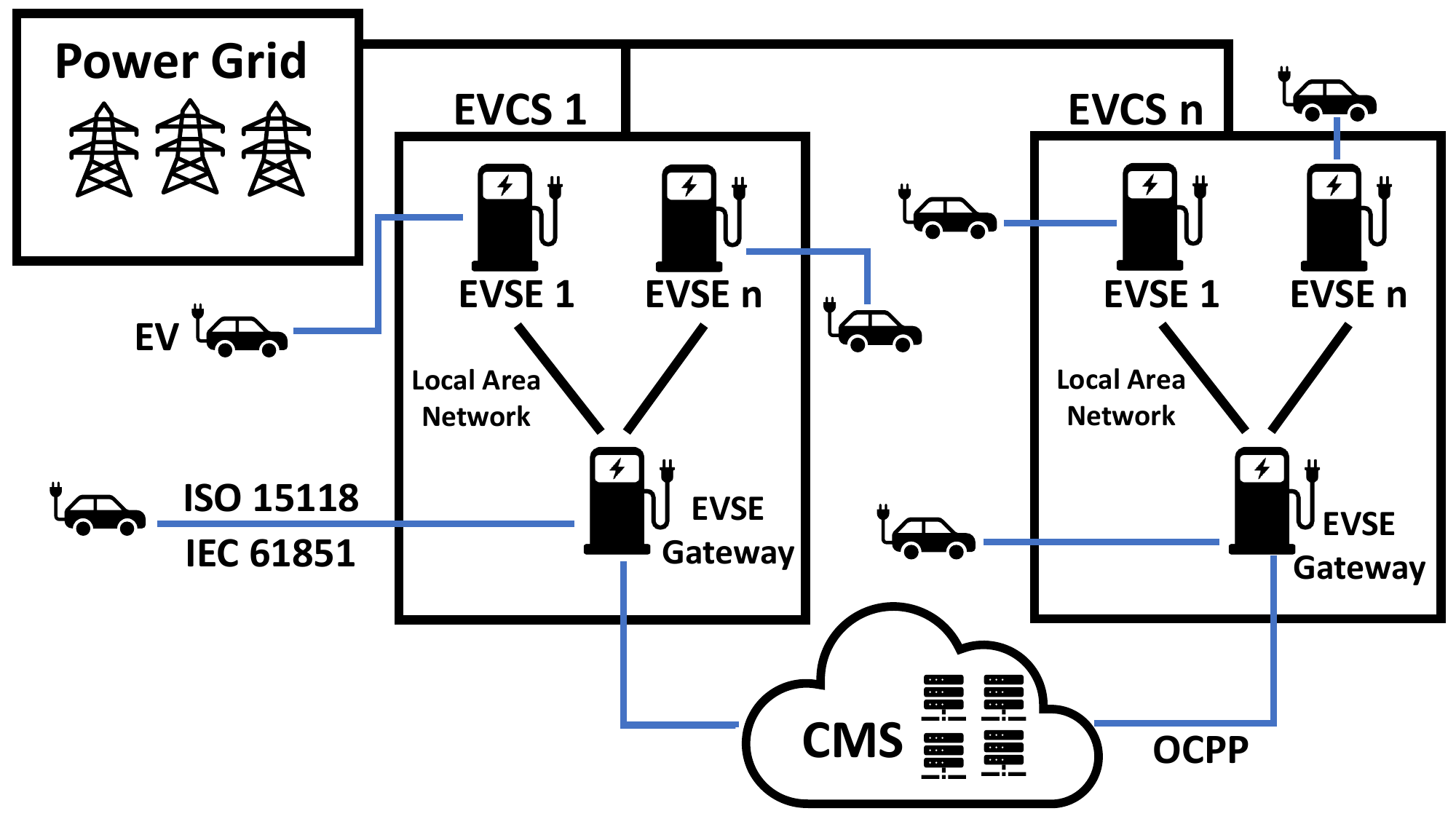}}
\caption{Electric vehicle infrastructure and protocols ~\cite{10.1145/3437258}}
\label{fig:fig5}
\end{figure}

The EVCI proposed by ElHussini et al.~\cite{10.1145/3437258} consists of four main entities: the EV, Power Grid, CMS, and EVCS. EVs, specifically Plug-In Hybrid Electric Vehicles (PHEVs) and Battery Electric Vehicles (BEVs), rely on charging stations for recharging. The Power Grid serves as the primary source of power for EVs through the charging stations. The EVCS encompasses the EVSE, which delivers power to EVs, and can be configured as a gateway or non-gateway. EVSEs vary in power output levels and can support unidirectional or bi-directional power flow. The EV-EVCS communication relies on standards such as ISO 15118 or IEC 61851, while the EVCS-CMS communication primarily utilizes the OCPP for standardized and efficient communication. Figure~\ref{fig:fig5} provides an overview of the key entities that constitute the infrastructure of the Electric Vehicle Charging System, along with the various protocols used for communication.

\subsection{Cyber Attack Scenarios in EVCI}



EVCSs provide power to EVs through an IoT-enabled infrastructure operating on specialized firmware. However, an attacker can disrupt the charging process if they gain access to the charging equipment via a Wi-Fi connection~\cite{hamdare2023cybersecurity}. Protecting EVCI from cyber threats is critical, as a wide range of potential attacks—including DoS, firmware manipulation, and malicious code injection—could disrupt charging operations, compromise user privacy, or destabilize the electric grid. Addressing these risks requires a comprehensive cybersecurity strategy that includes intrusion detection, standardized security practices, and robust authentication mechanisms~\cite{johnson2022cybersecurity}.

Cybersecurity attacks on EVCI can manifest in various forms. Since the communication system serves as the backbone of EVCI, handling crucial operations such as EV scheduling, authentication, and grid integration, any compromise in this communication can leave the entire EVCI vulnerable~\cite{basnet2021exploring}. Different motivations exist for attackers, ranging from electricity theft to disrupting a network of charging stations~\cite{gottumukkala2019cyber}.

The interconnected nature of EVs and charging stations presents significant challenges in terms of security and reliability. EVCIs are exposed to a wide range of cybersecurity threats, such as DoS attacks, MitM attacks, eavesdropping, spoofing, False Data Injection (FDI), hacking, ransomware, Trojan viruses, malware injection, and data breaches. These threats can lead to personal or financial data breaches and hijacking of power or charging infrastructure~\cite{hamdare2023cybersecurity, shirvani2023evaluation, girdhar2021hidden}.

Constant communication between EVSE and smart grids means that weaknesses in charging stations can have far-reaching consequences for the entire grid. The critical exchange of information occurs during vehicle charging or when users interact with their vehicles via software, creating a potential bottleneck where sensitive data and user privacy could be compromised. Securing this interconnected infrastructure is crucial for maintaining user privacy and ensuring reliable grid operations~\cite{shirvani2023evaluation}.

At the high-power, extreme-fast charging level, vulnerabilities are even more severe due to significant power flow. Attacks at this level can cause a loss of EV charging availability or safety concerns due to battery overcharging. Additionally, attackers could manipulate charging rates, block charging sessions, or even drain batteries, posing significant risks to both EV owners and charging station operators~\cite{girdhar2021hidden}.

One prominent attack is the DoS, where an attacker overwhelms the CMS or EVSE with excessive traffic or high charging requests, causing system resources to crash or become unresponsive~\cite{gottumukkala2019cyber, basnet2021exploring, hamdare2023cybersecurity}. DoS attacks can be initiated in various ways, such as through botnets or flooding. In research by Manoj and Mohd, hackers used spoofed IPs of legitimate EVSEs to launch DoS attacks~\cite{basnet2021exploring}. Another potential form of a DoS attack involves flooding the CAN bus of an EV with high volumes of traffic or spoofed messages, which could saturate the bus and disrupt the vehicle’s ability to perform normal operations, thereby preventing the vehicle from processing legitimate control messages. Additionally, if the communication link between the EV and the charging stations is flooded with excessive connection requests or malformed packets, the vehicle’s communication capabilities can be severely disrupted.

Gottumukkala et al.\cite{gottumukkala2019cyber} mention that a DoS attack on an EVCS often begins with a spoofing attack, where an attacker compromises a device's unique identifier, such as a MAC address, allowing them to masquerade as a legitimate user. Once the attacker gains unauthorized access to the system, they target the CMS or EVSE by flooding the network with excessive traffic or high charging requests. Attack variants like UDP or TCP/IP floods, low-rate DoS, ping floods, or Internet Control Message Protocol (ICMP) floods can take down a charging station or other nodes within the charging station network. This attack overwhelms system resources, causing the CMS to crash or become unresponsive, effectively preventing legitimate users from accessing charging services. To mitigate such attacks, EVCS operators should implement robust authentication mechanisms, traffic filtering, rate limiting, and anomaly detection techniques to identify and prevent abnormal traffic patterns\cite{gottumukkala2019cyber}.

Firmware manipulation is another significant threat, where attackers modify the charging station's firmware to inject malicious code, potentially affecting the processes, control actions, and operating conditions of the system. This malicious code can take the form of viruses, ransomware, or SQL and XSS injection attacks~\cite{basnet2021exploring}. Additionally, attackers can exploit EV connectors to install malware or manipulate charging settings, providing unauthorized access to the EVCS~\cite{hamdare2023cybersecurity}.

A lack of robust security in software modules makes charging stations and the cloud vulnerable to malware, enabling coordinated attacks that could shut down an entire network or impact the power grid by simultaneously activating multiple charging stations~\cite{gottumukkala2019cyber}. A more sophisticated attack, known as the Malicious Mode Attack (MMA), proposed by Weidong et al., targets the Coordinated Charging Load Control System (CCLCS) by sending forced oscillation control commands to multiple charging stations. This attack leads to high-amplitude forced oscillations in the power system by exploiting the weakly damped wide-area electromechanical mode, resulting in system instability and disrupting the power supply capacity of EVCSs~\cite{liu2022power}.

The OCPP, one of the most widely used protocols in charging stations, lacks built-in security measures against cyberattacks, potentially providing unauthorized users with an access point and leaving vulnerabilities for MitM attacks. An intruder can exploit the minimum status duration function in OCPP to access the communication interface between the EVCS and the CMS. Once inside, the attacker manipulates EVs and charge points, altering basic start or stop transaction functions to overcharge batteries or prevent them from charging altogether, potentially creating hazardous situations or inconveniencing consumers who need to charge later~\cite{gandhi2022impact}.

MitM attacks are common in radio-based communications between EVSEs, Plug-In Electrical Vehicles (PEVs), and Building Energy Management Systems (BEMS). An attacker can corrupt data or take full control of a node, relaying incorrect information such as false charging station status~\cite{gottumukkala2019cyber}. Moreover, MitM attacks can occur during the authentication and authorization process between PEVs and charging stations or throughout the charging session. Despite following the ISO 15118 security standard, the communication channel is still susceptible to MitM attacks. In these attacks, an intruder intercepts and manipulates the data exchanged between the charging station and the vehicle. By masquerading as a legitimate participant, the attacker can gain unauthorized access, potentially resulting in vehicle, information, or energy theft~\cite{shirvani2023evaluation}.

In an attack scenario proposed by ElHussini et al.~\cite{10.1145/3437258}, an adversary could manipulate an EVCS to cause a traffic bottleneck by altering the charging schedules of EVs. The attack unfolds in several steps: First, the attacker locates the EVCS using tools such as Shodan and Censys, gaining access to IP addresses and ports. Then, by exploiting security weaknesses such as default usernames and passwords found in manufacturers' manuals, the attacker takes control of the EVCS. This control enables the manipulation of EV charging schedules, redirecting charging traffic to specific locations and times, ultimately creating a bottleneck in the power grid. Finally, the attacker triggers disruptions at the grid level by coordinating sudden surges in power demand or reversing power flows from EVs back to the grid. These orchestrated attacks can lead to frequency instability, progressive failures, and imbalances between power demand and supply.

There are several other attacks, such as FDI, where attackers manipulate data transmission between EVSEs and management systems, leading to overcharging batteries or disrupting grid stability~\cite{hamdare2023cybersecurity}. Malware injections target publicly accessible EVSE units, introducing malicious software that can compromise the entire EVCS ecosystem, including the CMS and the power grid~\cite{hamdare2023cybersecurity}.

Additionally, poor database implementation can be exploited to insert, update, or delete data in the database through SQL injection attacks. This vulnerability enables attackers to inject malicious SQL commands to manipulate the database. They can execute commands affecting users' ability to charge, modify charging station location data, or alter station availability status, potentially causing public safety issues~\cite{gottumukkala2019cyber}.

To mitigate these risks, it is crucial to implement comprehensive cybersecurity measures in EVCI. The STRIDE framework helps identify potential security risks and vulnerabilities within EVCSs early in development. The six threats outlined in STRIDE include Spoofing, where attackers impersonate authorized sources to install malware on chargers, potentially causing station shutdowns; Tampering, involving the injection of false charging data that can mislead grid operations or damage EVs; Repudiation, where compromised chargers deny processing a charging fee, even if the EV owner has already paid for it, causing a financial loss to the EV owner; Information Disclosure, exposing sensitive communication between chargers and vehicles to unauthorized interception; DoS attacks that can flood the network and disable charging services; and Elevation of Privilege, where attackers gain unauthorized access to critical data and modify system files to disrupt the entire charging network. By using STRIDE, developers can proactively identify and address these threats, improving the safety and reliability of EVCI~\cite{girdhar2022machine}.


\subsection{TinyML: A Key Countermeasure in Cybersecurity for EVCI}


There are several methods to counter the attacks in the EVCI. The first line of defense involves securing networks with firewalls and filtering connections from trusted sources, alongside implementing strong authentication protocols. These measures are designed to prevent common exploits and protect against basic attempts at unauthorized access. Following these initial steps, an additional layer of security can be introduced to monitor for unusual activity in how the EVs are scheduled to charge. IDSs play a crucial role in this aspect, as they continuously monitor data traffic and user interactions, such as charging schedules, quickly identifying and responding to any deviations from normal patterns~\cite{10.1145/3437258}.

IDSs utilize various approaches based on the source of information and analysis strategies to safeguard against cyber threats. Host-based IDSs are deployed on individual hosts, using system, application logs, and audit trails to monitor and respond to malicious activities directly affecting the system. These IDSs focus on detecting unauthorized access, file tampering, and resource misuse by analyzing data from firewalls, routers, and operating systems. Conversely, network-based IDSs monitor the entire network's traffic to identify potential threats affecting multiple systems, relying on network sensors to manage the vast amounts of data for efficient threat detection~\cite{thakkar2022survey}.

Although network-based IDSs implemented at charging stations play a crucial role, this paper primarily focuses on enhancing security from within EVs themselves. Integrating host-based IDSs directly within EVs provides critical defense against specific threats at the point of interaction between the EV and the charging station, such as FDI and spoofing. By equipping EVs with robust host-based IDS, a comprehensive defense mechanism is established that not only protects the vehicle's systems but also secures data interactions during the charging process. This approach is pivotal in maintaining the integrity and security of the EV within the broader ecosystem of EVCI.

Considering that energy consumption and battery life are significant concerns in EVs, TinyML emerges as an effective tool for implementing IDS within these vehicles. A key advantage of TinyML is its ability to conduct real-time anomaly detection directly on the microcontrollers embedded within EVs. By processing data locally on these devices, the risk of data interception and unauthorized access is significantly reduced. This localized approach not only enhances security but also aligns with the sustainability goals of reducing overall energy usage in electric vehicles.

For this purpose, an ML model is aimed to be trained in a computationally robust environment, specifically a CMS, and then optimized and deployed for inference on an edge device, which in this case refers to a simple microcontroller in an EV. This approach draws from multiple methodologies, including, but not limited to, the Edge Learning Machine (ELM) framework. In this framework, models are trained in a desktop environment and subsequently utilized for predictions on STM microcontrollers~\cite{s20092638}.

The microcontroller, designed for IDS purposes, is installed within the EV to monitor the CAN bus. This monitoring is crucial as the CAN bus serves as the pathway for all operational data, managing various functions, including interactions with an EVCS~\cite{bozdal2020evaluation}. By examining traffic and communications for anomalies, the system identifies potential attacks that could bypass traditional security measures such as firewalls. It detects unusual or malicious activities, including those from compromised charging stations or incorrect charging data, which could compromise the vehicle's performance.

\begin{table*}[]
\centering
\caption{Label Distribution Before and After Merging}
\renewcommand{\arraystretch}{1.25} 
\begin{tabular}{|c|c|c|c|c|c|}
\hline
\textbf{Labels} & \textbf{Value Counts} & \textbf{Percentages} & \textbf{Merged Labels} & \textbf{Merged Value Counts} & \textbf{Merged Percentages} \\ \hline
BENIGN & 2,272,688 & 80.3245 & BENIGN & 2,272,688 & 80.3245 \\ \hline
DoS Hulk & 230,124 & 8.1334 & \multirow{4}{*}{Dos} & \multirow{4}{*}{251,712} & \multirow{4}{*}{8.8964} \\ \cline{1-3}
DoS GoldenEye & 10,293 & 0.3638 &  &  &  \\ \cline{1-3}
DoS slowloris & 5,796 & 0.2049 &  &  &  \\ \cline{1-3}
DoS Slowhttptest & 5,499 & 0.1944 &  &  &  \\ \hline
PortScan & 158,930 & 5.6171 & PortScan & 158,930 & 5.6171 \\ \hline
DDoS & 128,027 & 4.5249 & DDoS & 128,027 & 4.5249 \\ \hline
FTP-Patator & 7,938 & 0.2806 & \multirow{2}{*}{Brute Force} & \multirow{2}{*}{13,835} & \multirow{2}{*}{0.4890} \\ \cline{1-3}
SSH-Patator & 5,897 & 0.2084 &  &  &  \\ \hline
Web Attack-Brute Force & 1,507 & 0.0533 & \multirow{3}{*}{Web Attack} & \multirow{3}{*}{2,180} & \multirow{3}{*}{0.0770} \\ \cline{1-3}
Web Attack-XSS & 652 & 0.0230 &  &  &  \\ \cline{1-3}
Web Attack-Sql Injection & 21 & 0.0007 &  &  &  \\ \hline
Bot & 1,966 & 0.0695 & \multirow{3}{*}{Bot/Infiltration/Heartbleed} & \multirow{3}{*}{2,013} & \multirow{3}{*}{0.0711} \\ \cline{1-3}
Infiltration & 36 & 0.0013 &  &  &  \\ \cline{1-3}
Heartbleed & 11 & 0.0004 &  &  &  \\ \hline
\end{tabular}
\label{tableDataset}
\end{table*}

\subsection{TinyML limitations in EVCI and potential mitigation strategies}

Incorporating TinyML into EVCI presents a promising solution for enhancing security. However, the deployment of TinyML in this context is not without its challenges. This section explores the inherent limitations of integrating TinyML within EVCI, ranging from computational constraints to issues of scalability and interoperability. It also discusses potential mitigation strategies to overcome these barriers.

\subsubsection{Computational Power} Computational power in EVs is constrained due to the need for energy conservation~\cite{wang2023review}. Although TinyML models are optimized, implementing them in EVs remains a significant challenge. To mitigate this challenge, an effective strategy is to use model optimization. Different pruning and quantization techniques~\cite{han2015deep} can reduce model size and computational requirements, improving the efficiency of ML models on resource-constrained devices.
\subsubsection{Real-Time Processing} The real-time nature of IoT-based EVCSs demands instantaneous data flow for efficient and secure charging operations~\cite{kilichev2024next}. To make TinyML real-time, techniques such as quantization-aware training (QAT), pruning, and post-training quantization (PTQ) can significantly reduce model size and computational requirements while maintaining high accuracy, thereby reducing delay.
\subsubsection{Memory Capacity} EVs have limited storage space, which restricts the deployment of larger models. One strategy to mitigate this issue is to use memory-efficient model architectures and techniques such as feature selection to minimize data input size.
\subsubsection{Implementation Limitations} Since microcontrollers equipped with TinyML models are integrated into the EV's architecture, they can only be implemented in newly manufactured vehicles and not in older models. As a result, older vehicles lack the necessary hardware infrastructure to \subsubsection{Threat Adaptability} Scalability and adaptability to evolving threats remain challenging due to the dynamic and multifaceted nature of the IoT-based environment of EVCI~\cite{kilichev2024next}. Continuous learning~\cite{ravaglia2021tinyml} is an effective technique to update TinyML models based on new attack data so that anomaly detection models can identify previously unseen threats.
\subsubsection{Standardization and Interoperability} The diversity in communication protocols used by charging stations (e.g., SAE, IEC 61851, ISO 15118, CHAdeMO) poses a challenge for developing a universal TinyML model. One strategy to address this issue is to design modular models that can adapt to different protocols through customizable preprocessing and feature extraction layers.  

\section{Experimental Setup and Results}

In this section, a Python-based simulation is presented, aiming to compare the performance of traditional ML methodologies against TinyML in the context of an IDS.

\subsection{Dataset Description}

The EVCI is vulnerable to a wide range of cyberattacks, making robust anomaly detection crucial. For this purpose, the CICIDS2017 dataset~\cite{Sharafaldin2018TowardGA, cicids2017} is selected, which, despite containing general attacks, aligns well with the types of threats found in EVCI. Attacks such as DoS or web attacks, which are included in the CICIDS2017 dataset, are also prevalent in electric vehicle charging infrastructures, as discussed in previous sections.

The CICIDS2017 dataset is provided by the Canadian Institute for Cybersecurity and serves as a foundational resource for cybersecurity research. In more detail, this dataset includes a diverse range of common cyber-attacks, including DoS (encompassing DoS Hulk, DoS GoldenEye, DoS slowloris, and DoS Slowhttptest), PortScan, DDoS, Brute Force (including FTP-Patator and SSH-Patator), Web Attack (comprising Web Attack-Brute Force, Web Attack-XSS, and Web Attack-SQL Injection), and a combined category of Bot/Infiltration/Heartbleed~\cite{panigrahi2018detailed}.

Using IoT-based datasets for anomaly detection in EVCI has been demonstrated in other studies as well. For instance, in a paper by Kilichev et al.\cite{kilichev2024next}, EVCSs are safeguarded using robust IDS. The authors employ a deep learning-based model evaluated on the Edge-IIoTset dataset\cite{ferrag2022edge}, a comprehensive real-world IoT security dataset, to enhance detection accuracy and optimize resource efficiency.

Overall, the CICIDS2017 dataset provides a set of realistic traffic scenarios, enabling comprehensive detection capabilities that improve the reliability, safety, and efficiency of EVCI, therefore selected for this research.


\subsection{EVCI Dataset Creation}
The dataset exhibits class imbalance, making it important to balance the chance of occurrence for each class label for more accurate analysis. To address this issue, specific minority classes are merged into larger, more representative ones~\cite{9013892, panigrahi2018detailed}. The details of these merged classes, including the original and new labels as well as their prevalence ratios, can be found in Table~\ref{tableDataset}.

Additionally, the dataset is quite large, containing over 3 million samples. While the large volume of data is advantageous for comprehensive analysis, it can also become computationally intensive. To mitigate this challenge, data sampling is proposed to reduce the dataset size to just 5\% of the total. To make sure the original distribution of classes is maintained in this smaller dataset, stratified sampling across different output classes is used. Therefore, the smaller dataset still gives us a comprehensive view of the original data's diverse labels.

The CICIDS2017 dataset comprises 85 distinct features. These features provide a comprehensive representation of network traffic, enabling the development of sophisticated ML models for IDS. However, not all these features would be relevant in the context of detecting cyber-attacks in EVCI communications. For our specific purpose, a subset of features are carefully selected which are more likely to exhibit patterns related to potential cyber attacks in the EV charging environment. The key features to consider would be:

\begin{table*}
\centering
\caption{Summary of Scenarios for Model Evaluation}
\renewcommand{\arraystretch}{1.25} 
\label{tab:ModelScenarios}
\small 
\setlength{\tabcolsep}{3pt} 
\begin{tabular}{|M{6cm}|M{8cm}|c|} 
\hline
 \textbf{Model Type / Optimization} & \textbf{Description} & \textbf{Designation} \\
\hline
 MLP / Traditional & Initial implementation of an MLP model & ML\_MLP \\
\hline
 MLP / Resource-Optimized & Converted MLP model using TFLite & TinyML\_MLP \\
\hline
 RF / Traditional & Initial implementation of an RF model & ML\_RF \\
\hline
 RF / Resource-Optimized & RF model with reduced computational footprint & TinyML\_RF \\
\hline
\end{tabular}
\end{table*}

\begin{itemize}
    \item Flow ID: Useful for uniquely identifying each network flow in the traffic, which would help in distinguishing between different sessions.
    \item Source IP: Identifies the origin of the network flow. Changes or unusual patterns in the source IP can potentially indicate an attack.
    \item Source Port: Similar to the Source IP, the source port can be indicative of the type of service being used, and observing unusual patterns in this can also help detect potential attacks.
    \item Destination IP: The IP address where the traffic is being sent. Unusual destination IPs may indicate a network attack.
    \item Destination Port: The port to which traffic is being sent. Different ports are used by different services, and a sudden change in destination port could be a sign of an attack.
    \item Protocol: The network protocol being used. A sudden shift from the normal protocol could indicate an attack.
    \item Timestamp: The timestamp of each network flow can be crucial in identifying temporal patterns in the traffic, which could be indicative of an attack.
    \item Flow Duration: Long durations might indicate attempts to keep a connection open for an attack.
    \item Total Fwd Packets and Total Backward Packets: Unusually large numbers of packets being sent could be indicative of a DoS attack.
    \item Flow Bytes/s and Flow Packets/s: The rate at which bytes and packets are transferred can be indicative of potential anomalies and cyber threats. 
    \item Flow IAT Mean, Flow IAT Std, Flow IAT Max, Flow IAT Min: These features provide statistics about the Inter-Arrival Times (IAT) of packets in a flow. Variations in these values can signify irregular network behavior.
    \item Fwd PSH Flags, Bwd PSH Flags: These flags indicate the use of the PSH function in TCP. Abnormal patterns can point to unusual network behavior.
    \item Fwd Header Length, Bwd Header Length: The length of packet headers can be indicative of anomalies, as some attack methods involve manipulating packet headers.
    \item FIN Flag Count, SYN Flag Count, RST Flag Count, PSH Flag Count, ACK Flag Count, URG Flag Count, CWE Flag Count, ECE Flag Count: Flags in network packets can often provide useful insights into the nature of the traffic. Unusual flag patterns can often signify network intrusions.
    \item Init\_Win\_bytes\_forward, Init\_Win\_bytes\_backward: The initial window size in TCP can sometimes be manipulated in certain types of network attacks.
\end{itemize}

These 30 features help in understanding both the volume and nature of the traffic passing through the network. Certain features require encoding to be effectively processed by ML algorithms. Specifically, the \texttt{Flow ID}, \texttt{Source IP}, \texttt{Destination IP}, and \texttt{Protocol} attributes are categorical in nature and therefore should be encoded. The \texttt{Label} attribute, which comprises 15 categories representing various types of attacks, will also need encoding as it is to be used as a target variable in a classification task. Encoding these features help transform non-numeric data into a numeric form that can be understood and used by our ML algorithms.

\subsection{Data Pre-Processing}

In the given dataset, all the timestamps fall within the same month and year. This particular aspect, therefore, does not contribute to any significant variance in the data that could impact the prediction models' performance. What might influence the model, though, are the patterns and trends related to the time of day and the day of the week when network events occur. Consequently, from the \texttt{Timestamp}, only two features - the day of the week and the hour - are extracted. Furthermore, the minutes and seconds are discarded from the analysis. While it might provide a more granular perspective on time, it also adds additional complexity to the data which leads to more computational overhead. Given the nature of the problem, where seconds and minutes are less likely to dramatically impact overall trends, the benefit of including them does not justify the additional processing cost. After extracting these two features, the dataset now includes a total of 31 features.

As a part of our preprocessing strategy, the dataset is scanned for the presence of any Not a Number (NaN) values. Any sample containing missing or NaN values is removed from the dataset. Furthermore, any positive infinite values found within the dataset have been replaced with the maximum finite representable positive value for a float32 data type. This strategy is implemented to overcome issues related to computational limitations of handling infinite values, which can potentially disrupt model training. 

Finally, twenty percent of the dataset has been allocated as the test set. Additionally, stratification is employed to guarantee that the distribution of attack categories in the training and test sets is the same as the original dataset. For normalizing feature values, the StandardScaler from the sklearn library is used.

\subsection{Model Description}

For the simulation, four distinct scenarios are considered to evaluate both traditional and resource-optimized ML models. Initially, a Multi-Layer Perceptron (MLP) model is implemented as a baseline for model simplicity, which is then converted to its resource-optimized counterpart using TFLite. These two models are referred to as ML\_MLP and TinyML\_MLP, respectively. Alongside this, Random Forest (RF) is selected as the traditional ML model for comparison. To create its resource-optimized version, specific metrics and parameters are modified to reduce its computational demand. These two RF models are designated as ML\_RF and TinyML\_RF. RF is selected for its proven high detection rate in identifying network attacks compared to other models~\cite{al2023machine}. A summary of the four distinct scenarios considered for model evaluation is provided in Table~\ref{tab:ModelScenarios}.

\begin{figure*}[!t]
\centerline{\includegraphics[scale=.65]{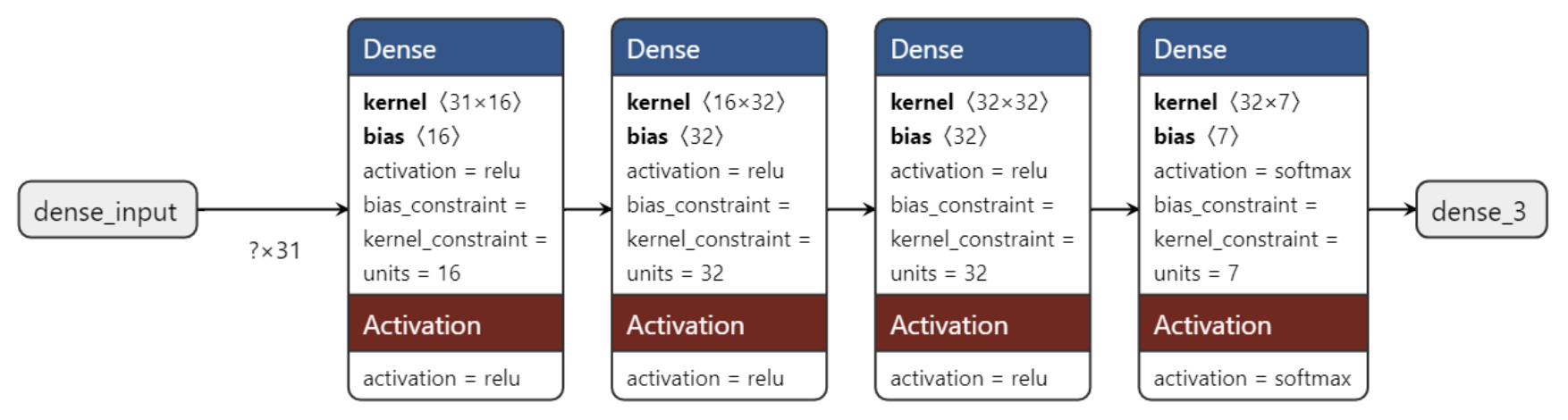}}
\caption{\centering Visual representation of ML\_MLP as analyzed by Netron~\cite{Netron}}
\label{ML_MLP}
\end{figure*}

\begin{figure*}[!t]
\centerline{\includegraphics[scale=.65]{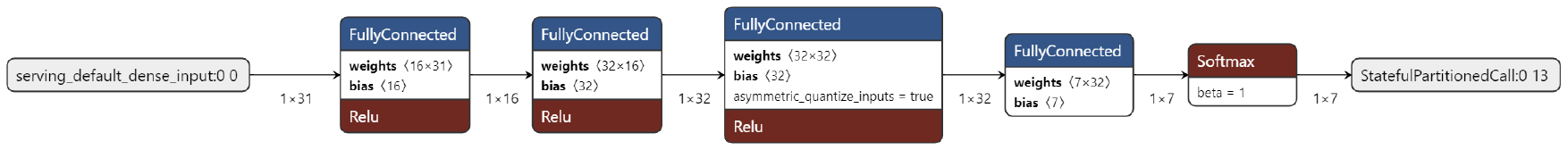}}
\caption{\centering Visual representation of TinyML\_MLP as analyzed by Netron~\cite{Netron}}
\label{TinyML_MLP}
\end{figure*} 

For ML\_MLP models, a sequential model is chosen from the TensorFlow Keras library. This model is constructed with one dense layer containing 16 units, followed by two dense layers, each containing 32 units, all using the ReLU activation function. The final layer is a dense layer with a size matching the number of unique classes in the training dataset, employing a softmax activation function to output probabilities for each class. The model is compiled with the Adam optimizer, using the sparse categorical cross-entropy loss function and accuracy as the evaluation metric. During training, an early stopping callback is utilized to monitor the validation loss, which helps prevent overfitting by stopping the training process if the validation loss does not improve for 10 consecutive epochs. The best model weights are restored upon early stopping. Additionally, a validation split of 10\% is used during training, wherein 10\% of the training data is set aside for validation purposes, and the model is trained on the remaining 90\% of the data. 

The TinyML\_MLP model is created by converting a TensorFlow model to TFLite format using default optimizations, which primarily involves quantization to reduce model size and speed up inference, however with a potential minor impact on accuracy. Due to not specifying optimization techniques during the conversion of the TensorFlow model to its TFLite counterpart, Netron, a tool that visually represents the architecture and settings of ML models, is employed to scrutinize which optimization methods are applied during the conversion~\cite{Netron}. Netron provides an intuitive graphical representation of the models' layers, activations, and other key attributes, making it easier to understand their internal structures and transformations through the optimization process. The visual representations of these models can be seen in Figures~\ref{ML_MLP} and~\ref{TinyML_MLP}.

\begin{table*}[h]
    \centering
    \small
    \caption{Average Computational Resources Across 5-Fold Cross-Validation (each containing around 28,000 Test Samples) and Model Size}
    \begin{tabular}{|c|c|c|c|}
        \hline
        \textbf{Model} & \textbf{Inference Time (ms)} & \textbf{Inference Memory Usage (KB)} & \textbf{Model Size (KB)} \\
        \hline
        ML\_MLP & 86.1553 & 117.7058 & 67.2813 \\ \hline
        TinyML\_MLP & 0.2400 & 0.9023 & 8.6953 \\ \hline
        ML\_RF & 17.3951 & 18.7856 & 2380.1885 \\ \hline
        TinyML\_RF & 3.1817 & 11.8481 & 134.5010 \\
        \hline
    \end{tabular}
    \label{tab:MergedResourcesAndSize}
\end{table*}

The comparison between the two neural network architectures from ML\_MLP to TinyML\_MLP reveals subtle but significant modifications. In the ML\_MLP architecture, the batch size is unspecified, as indicated by the question mark; this represents a flexible batch size that can be set dynamically. In contrast, the TinyML version specifies a single input, suggesting a fixed batch size of one, which is typical for TinyML applications where the model operates on a single data point at a time for real-time or near-real-time inference on edge devices. The last layer of the TinyML model architecture shows a notable alteration; it lacks an activation function within the layer itself, in contrast to the ML\_MLP model where the softmax activation is integrated within the final Dense layer. Instead, the TinyML architecture appends an explicit separate Softmax layer, which is a common practice to provide more control over the application of this activation, especially in quantized models where numerical precision is of high concern. The beta parameter set to 1 for the Softmax layer indicates that the function's output is not scaled or modified beyond the standard Softmax behavior. The observation of the attribute asymmetric\_quantize\_inputs = true in the TinyML model is indicative of quantization, a process that converts the model from floating-point numbers to lower-precision representation, reducing model size and potentially increasing computational efficiency on hardware that favors integer operations. However, it should be noted that Netron visualizes the static structure of the model but does not capture dynamic runtime behaviors or optimizations that occur during actual model execution, such as memory management techniques, dynamic quantization, and hardware-specific accelerations. Notably,  the model visualizations remained consistent across all five folds.

The ML\_RF model uses default settings with 100 decision trees, no maximum tree depth, and a random state of 0 for reproducibility, while for feature selection, features are sorted by importance and cumulatively added till a 60\% importance threshold is reached. This feature selection method allows the following TinyML\_RF model to be trained only on the most impactful variables. 

Subsequently, the TinyML\_RF model is created by reducing the decision trees to 10, capping tree depth at 10, with a consistent random state of 0 for reproducibility. Meanwhile, the model's complexity is simplified by training only on the previously selected impactful variables.

The prediction phase measures time and memory usage per sample. Additionally, accuracy, precision, recall, and F1-score, are calculated to demonstrate the model's overall predictive ability.

Computations were performed on a system equipped with an Intel(R) Core(TM) i7-9700 CPU, which operates at a base frequency of 3.00 GHz. The system is configured with 32.0 GB of installed RAM, of which 31.8 GB is usable, and runs on a 64-bit Windows operating system with an x64-based processor architecture.

\begin{table*}
    \centering
    \small
    \caption{Average Performance Metrics Across 5-Fold Cross-Validation (each containing around 28,000 Test Samples)}
    \begin{tabular}{|c|c|c|c|c|}
        \hline
        \textbf{Model} & \textbf{Accuracy (\%)} & \textbf{Precision (\%)}  & \textbf{Recall (\%)} & \textbf{F1 Score (\%)} \\
        \hline
        ML\_MLP & 99.9724 & 99.9729 & 99.9724 & 99.9711 \\ \hline
        TinyML\_MLP & 99.9724 & 99.9729 & 99.9724 & 99.9711 \\ \hline
        ML\_RF & 99.9880 & 99.9880 & 99.9880 & 99.9878 \\ \hline
        TinyML\_RF & 99.9180 & 99.9201 & 99.9180 & 99.9179 \\
        \hline
    \end{tabular}
    \label{tab:AccuracyMetrics}
\end{table*}

\subsection{Results and Analysis}

In the analysis, a 5-fold cross-validation strategy is employed to ensure robustness and mitigate the impact of dataset-specific randomness on the evaluation metrics. Each fold contains a distinct 5\% subset of the overall dataset and is generated using stratified sampling to maintain the original class distribution. This approach enables a more reliable and less biased representation of the model's performance to be obtained. The metrics and evaluations discussed in this section are derived from the mean values across these five folds.

\begin{table*}[h]
    \centering
    \caption{Performance Metrics for Enhanced MLP Architectures}
    \renewcommand{\arraystretch}{1.25} 
    \begin{tabular}{|c|c|c|c|c|}
        \hline
        \textbf{Model} & \textbf{Accuracy (\%)} & \textbf{Precision (\%)}  & \textbf{Recall (\%)} & \textbf{F1 Score (\%)} \\
        \hline
        ML\_MLP & 99.8162 & 99.7564 & 99.8162 & 99.7848 \\ \hline
        TinyML\_MLP & 99.8127 & 99.7529 & 99.8127 & 99.7813 \\
        \hline
    \end{tabular}
    \label{tab:MLPPerformanceMetrics}
\end{table*}

Table~\ref{tab:MergedResourcesAndSize} presents a comparative analysis of the computational efficiency and model sizes of different ML models—specifically, MLPs and RFs—in both their traditional (ML) and resource-constrained (TinyML) configurations. Notably, the TinyML versions exhibit a substantial reduction in both prediction time and memory usage. For example, TinyML\_MLP operates nearly 360 times faster and requires significantly less memory (about 130 times less) than its ML\_MLP counterpart. Similarly, TinyML\_RF also demonstrates remarkable efficiency, taking less time (approximately 5.5 times faster) and consuming significantly less memory (around 1.6 times less) than ML\_RF. The conversion of the MLP model to its TFLite variant, TinyML\_MLP, is notably efficient, taking around 1.2 seconds on average across five folds. Additionally, the table reveals the sharp contrast in model sizes of various configurations. The model size of the resource-optimized MLP is reduced to around 9 KB, almost an eighth of its resource-intensive counterpart at around 70 KB. Even more noticeable is the difference in the RF models; the resource-optimized version has a model size of just about 135 KB, a dramatic reduction from the resource-intensive version's substantial 2.32 MB.

Table~\ref{tab:AccuracyMetrics} presents a thorough comparison of key performance metrics —accuracy, precision, recall, and F1 score— across the same models in both resource-intensive and resource-optimized configurations. Remarkably, the resource optimizations in TinyML variants have little to no impact on their predictive performance. For instance, TinyML\_MLP achieves the same metrics as its ML counterpart across all evaluated parameters. Even for TinyML\_RF, where a minor drop in performance can be observed, the metrics remain impressively high which indicates a marginal trade-off for the significant savings in computational resources.

The remarkable efficiency in both time and memory usage for TinyML\_MLP and TinyML\_RF, with maintained or nearly similar accuracy to their traditional counterparts, can be attributed to proper optimizations for TinyML environments. For TinyML\_MLP, the key lies in the model compression technique, especially quantization, which reduces the precision of the weights from floating-point to lower-bit representations. This process significantly reduces the model's size and speeds up computation, enabling TinyML\_MLP to maintain high accuracy while operating much faster and with less memory on constrained devices. For TinyML\_RF, the efficiency improvement comes from model simplification and tree pruning. By reducing the depth and number of trees and optimizing the decision nodes, the model size and required computational resources are significantly decreased. This streamlined version of the RF model still captures the essential patterns in the data, enabling a substantial reduction in inference time and memory usage with a minimal impact on accuracy. These techniques illustrate TinyML's capacity to improve efficiency on devices with limited resources.

\subsection{Visualization} 

\begin{figure*}[htp]
    \centering
    
    \begin{subfigure}{0.49\textwidth}
        \centering
        \includegraphics[width=\textwidth]{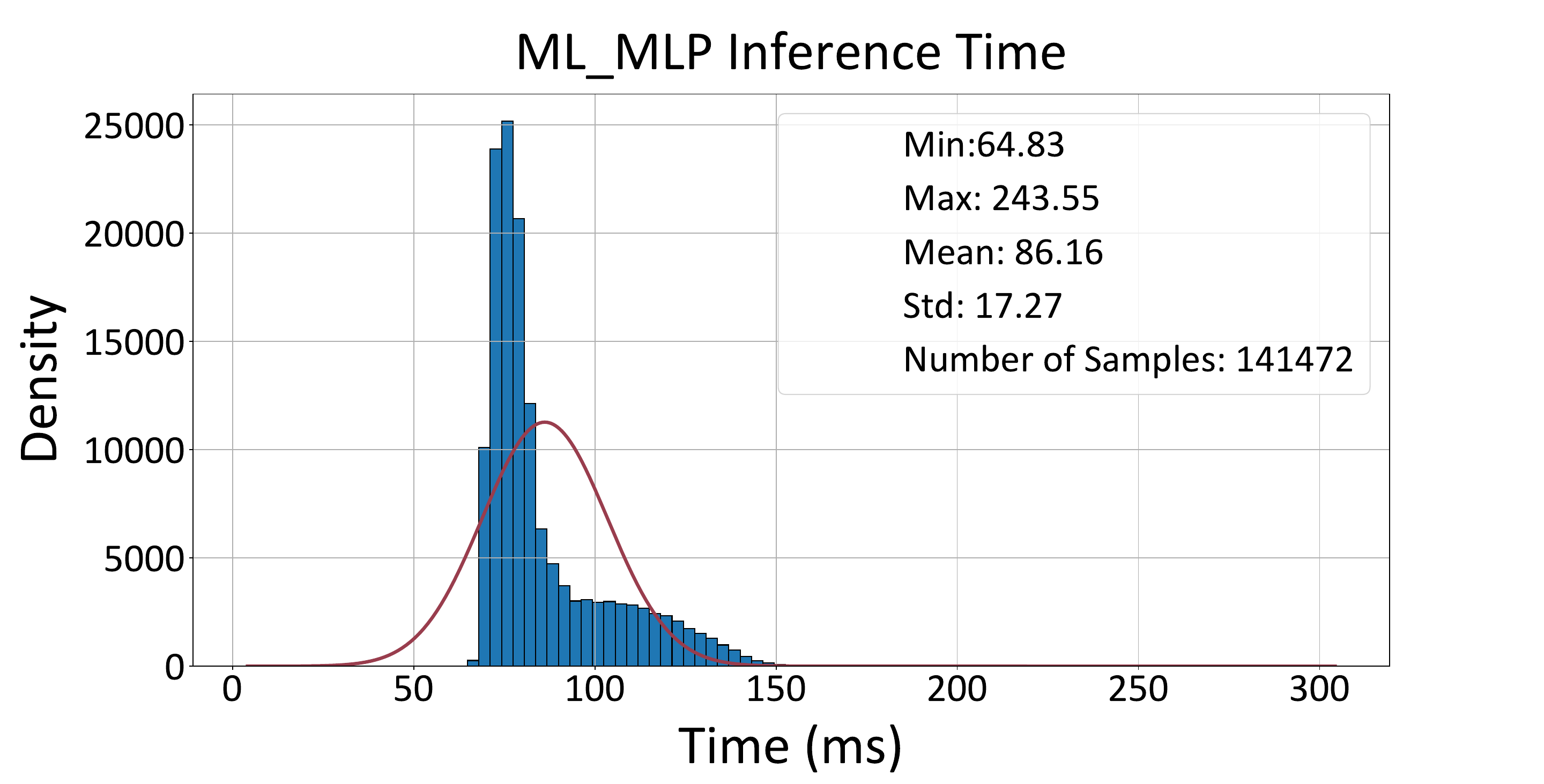}
        \label{fig:sub1}
    \end{subfigure}
    \begin{subfigure}{0.49\textwidth}
        \centering
        \includegraphics[width=\textwidth]{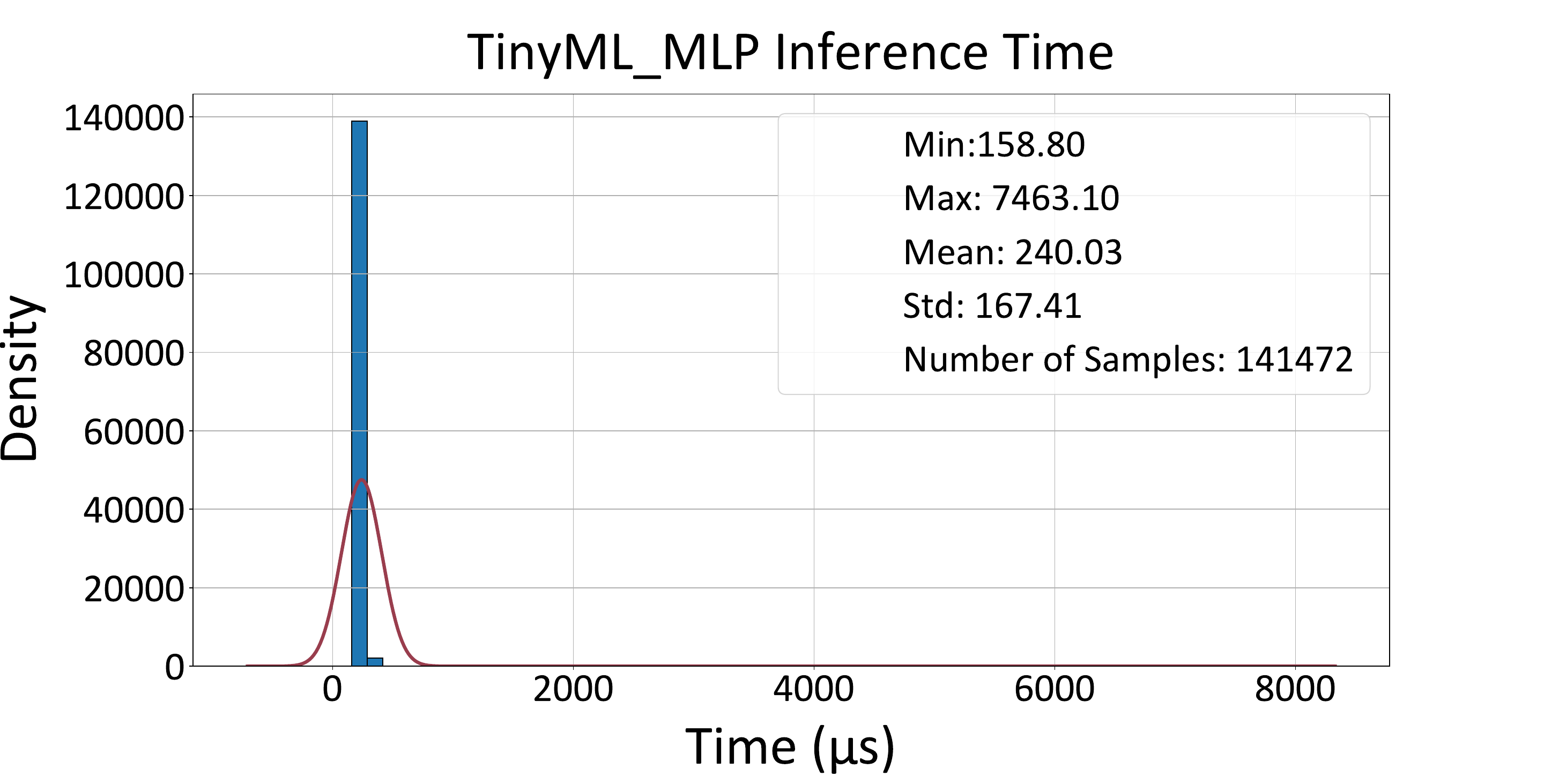}
        \label{fig:sub2}
    \end{subfigure}
    
    
    \begin{subfigure}{0.49\textwidth}
        \centering
        \includegraphics[width=\textwidth]{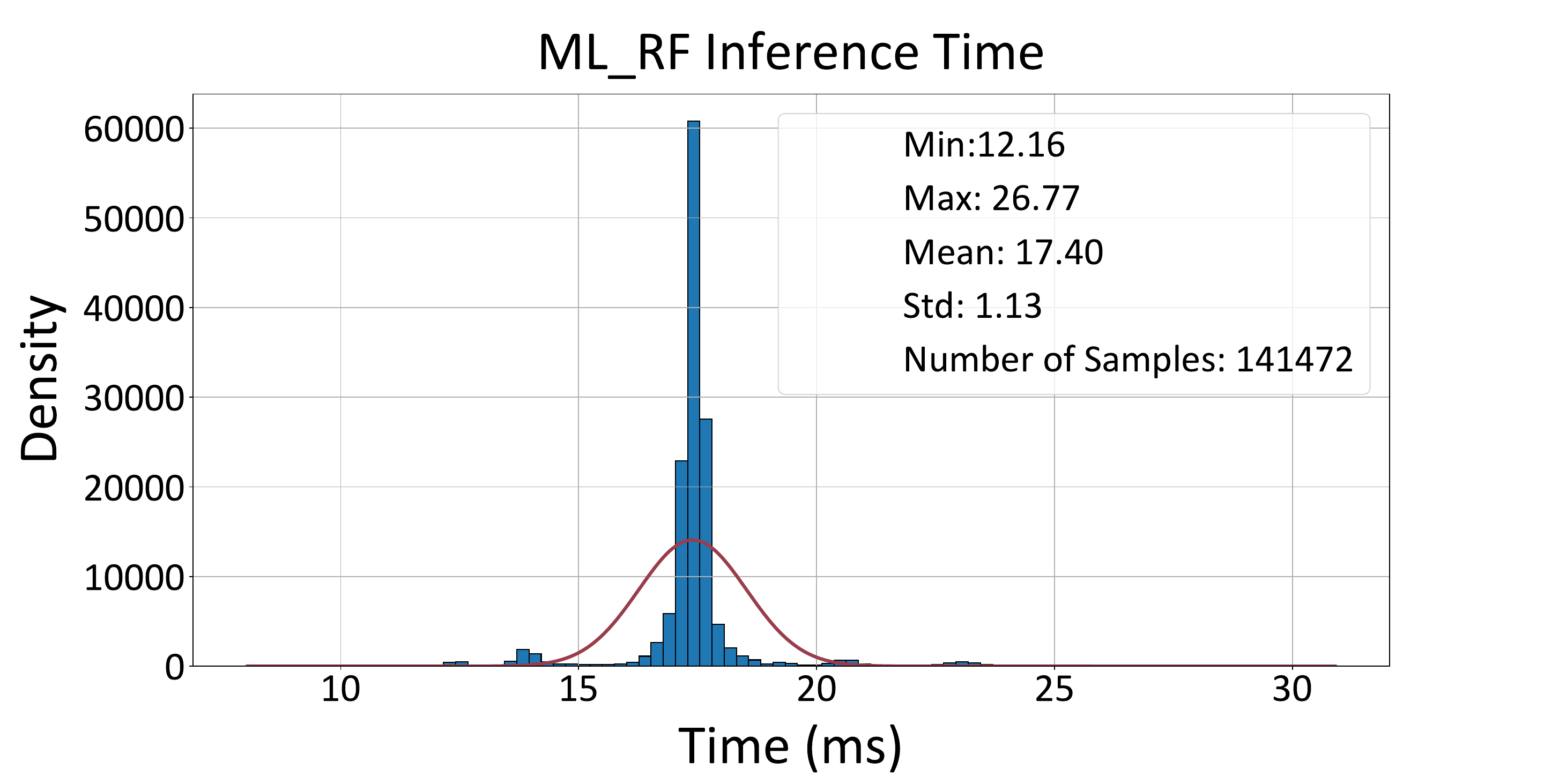}
        \label{fig:sub3}
    \end{subfigure}
    \begin{subfigure}{0.49\textwidth}
        \centering
        \includegraphics[width=\textwidth]{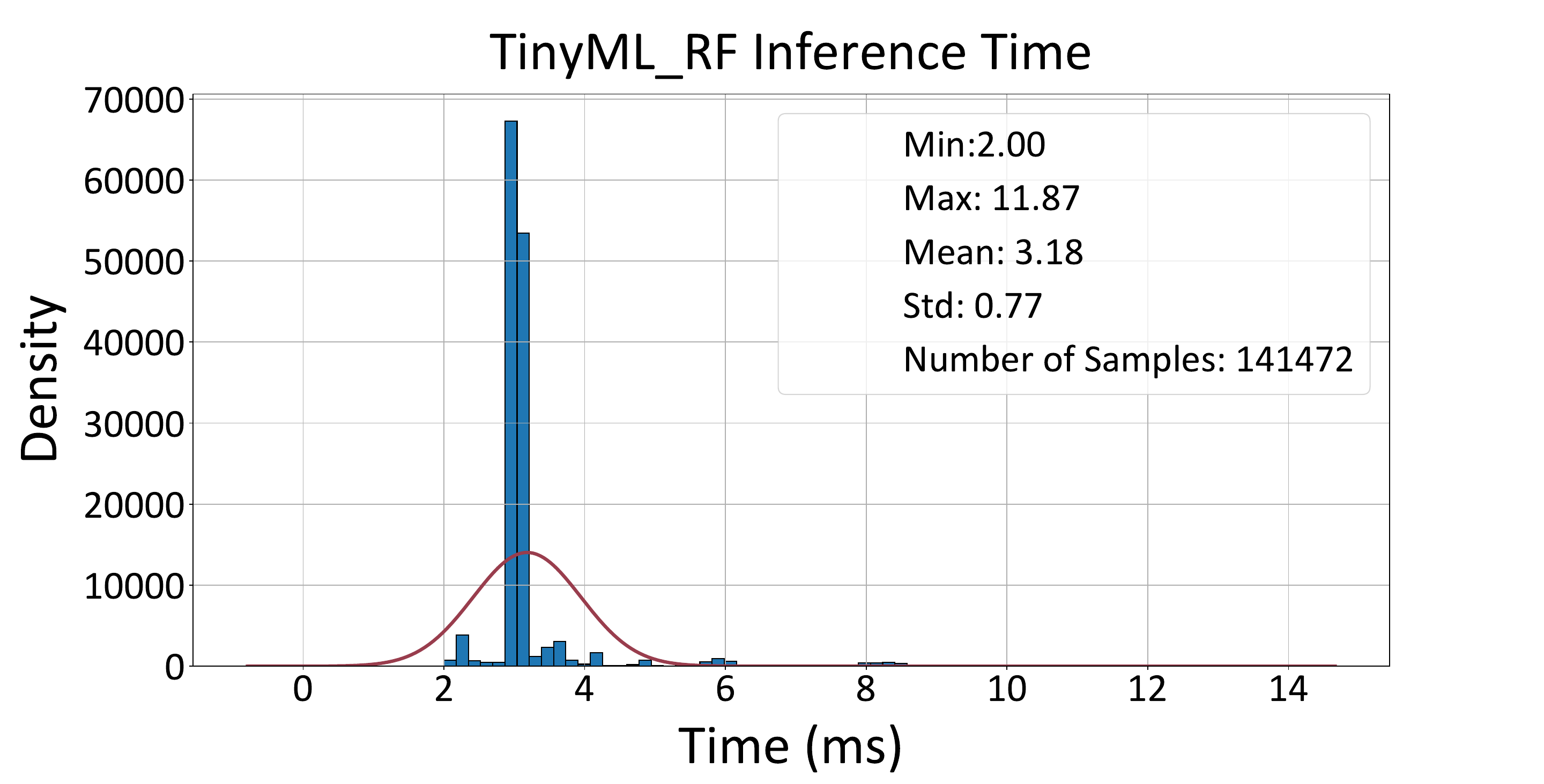}
        \label{fig:sub4}
    \end{subfigure}
    \caption{Histogram representation of per data-sample inference time during prediction}
    \label{fig:main2}
\end{figure*}

\begin{figure*}[htp]
    \centering
    
    \begin{subfigure}{0.49\textwidth}
        \centering
        \includegraphics[width=\textwidth]{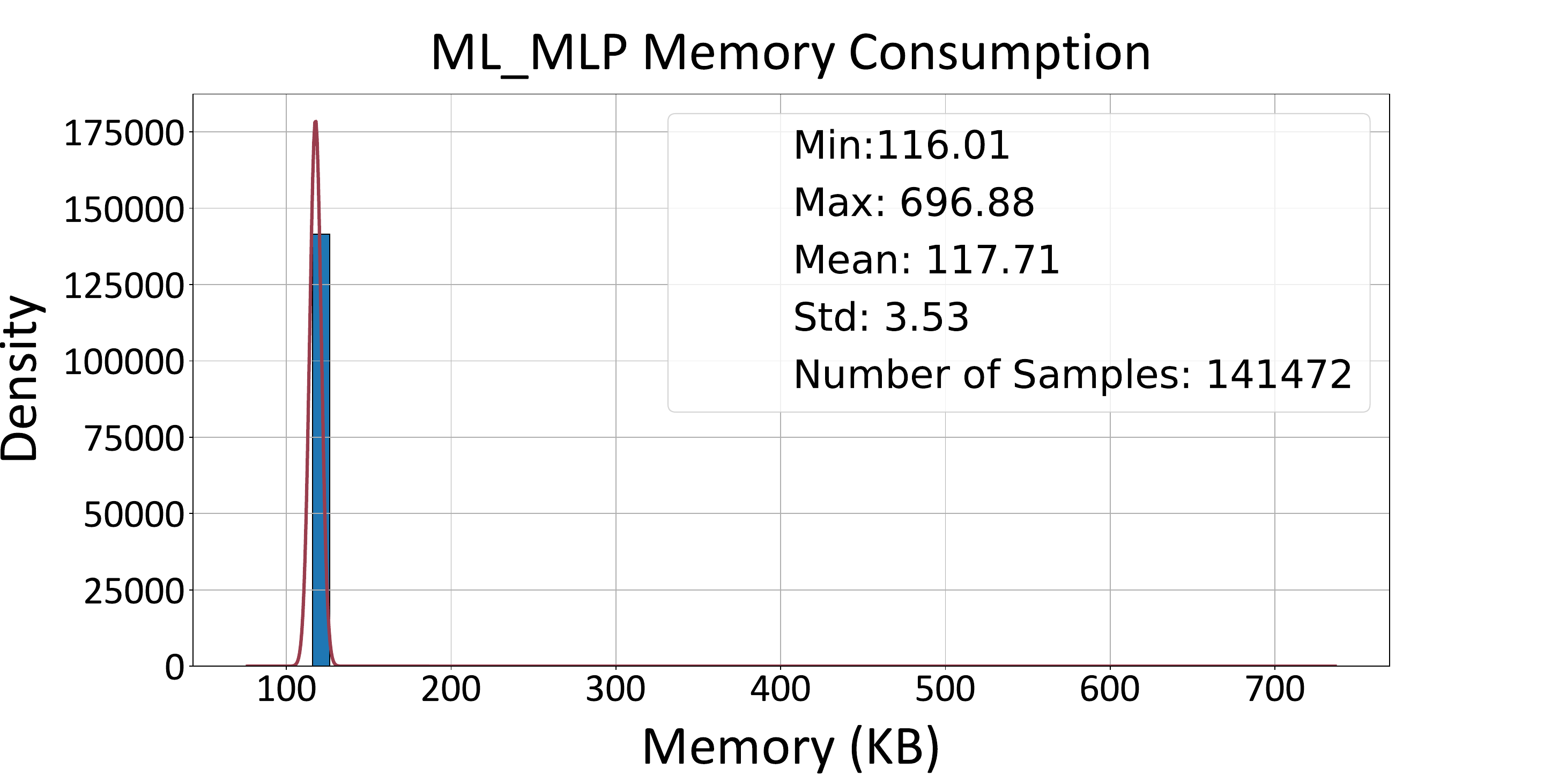}
        \label{fig:sub1}
    \end{subfigure}
    \begin{subfigure}{0.49\textwidth}
        \centering
        \includegraphics[width=\textwidth]{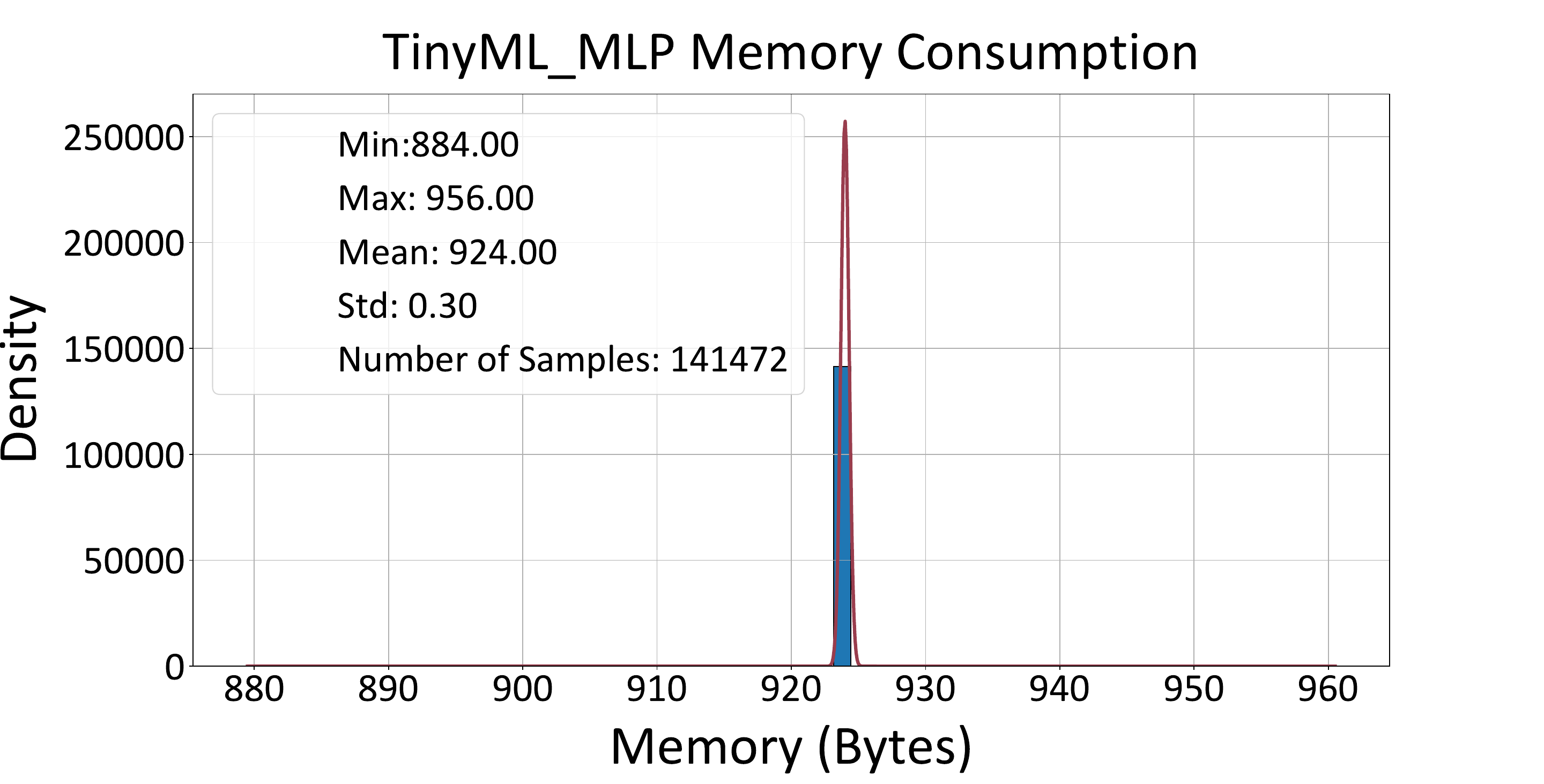}
        \label{fig:sub2}
    \end{subfigure}
    
    
    \begin{subfigure}{0.49\textwidth}
        \centering
        \includegraphics[width=\textwidth]{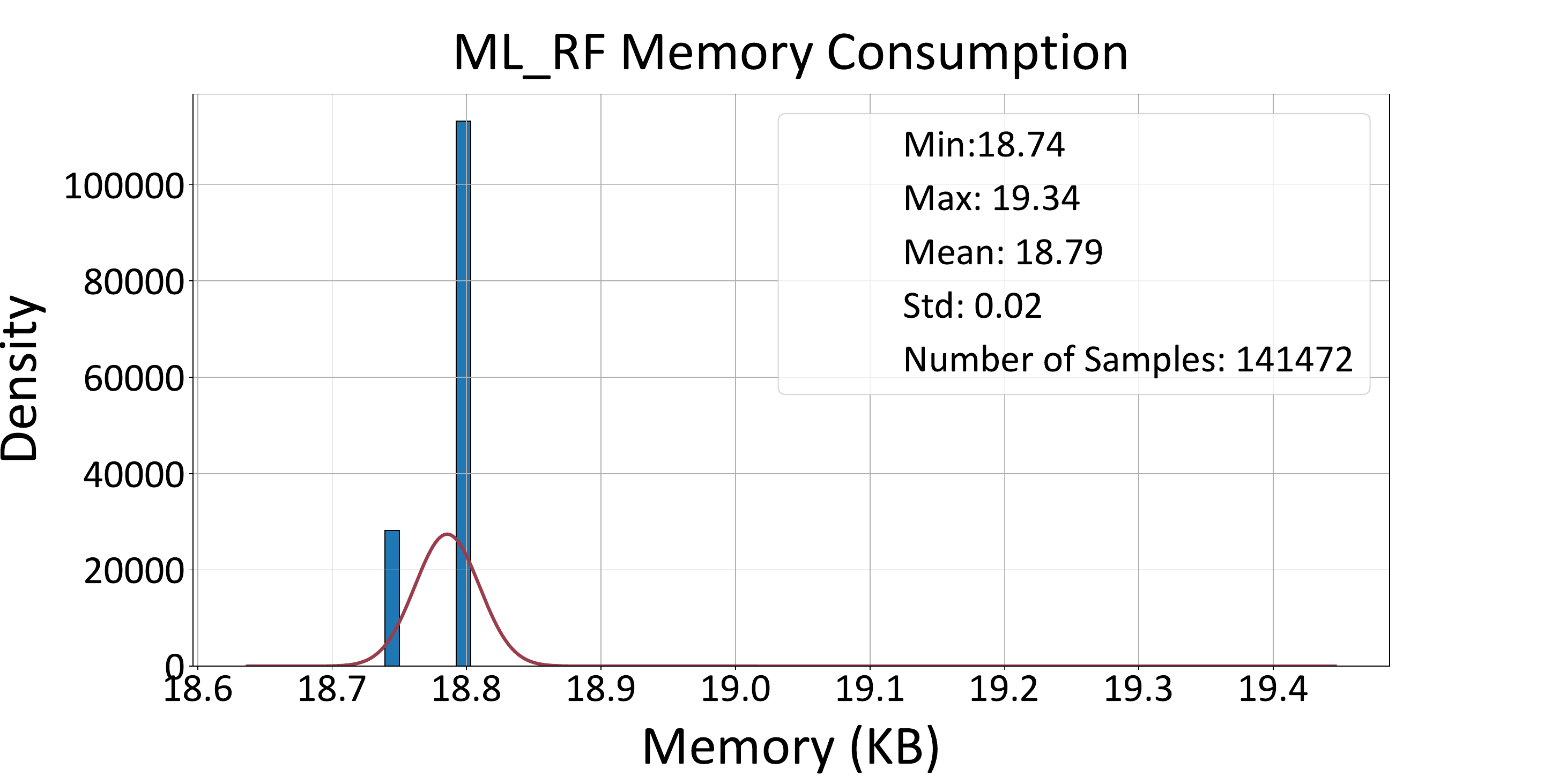}
        \label{fig:sub3}
    \end{subfigure}
    \begin{subfigure}{0.49\textwidth}
        \centering
        \includegraphics[width=\textwidth]{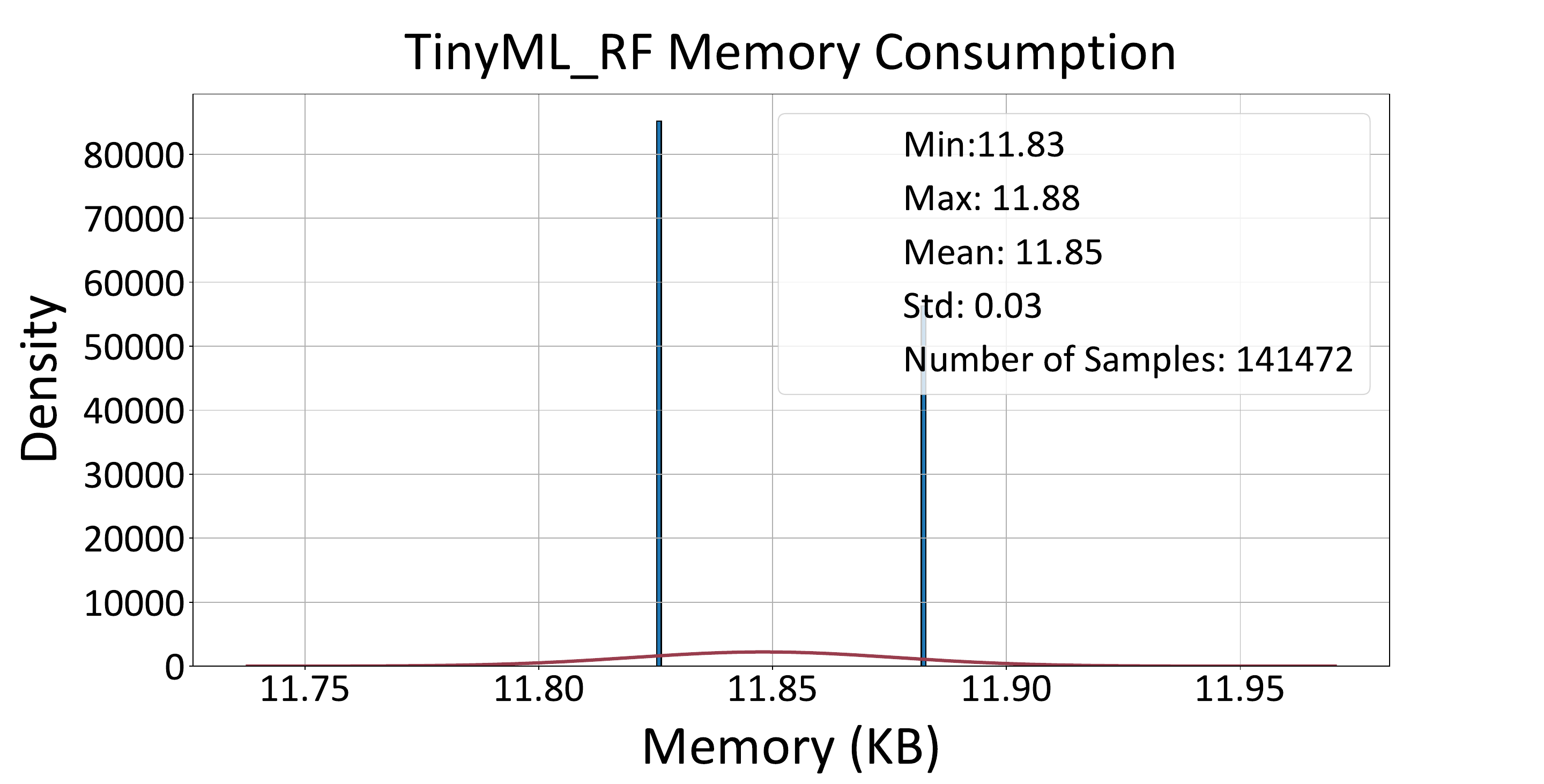}
        \label{fig:sub4}
    \end{subfigure}
    \caption{Histogram representation of per data-sample inference memory usage during prediction}
    \label{fig:main1}
\end{figure*}

For a deeper comprehension of the prediction phase's behavior, histogram plots are provided in the analysis. The histogram presents a combined representation of prediction samples from five folds. Figure~\ref{fig:main2} displays the inference time for each method and algorithm, while Figure~\ref{fig:main1} provides insight into their memory usage. Relevant statistics, such as the minimum, maximum, mean, standard deviation, and the total count of combined samples, are included as a legend within the figure for clarity. Moreover, a normal distribution curve has been applied on the histogram to offer a comparative baseline. This curve has been adjusted by scaling it with the product of the number of samples and the bin width to align appropriately with the histogram's distribution.

\subsection{TinyML Robustness}

In initial experiments, both TinyML and traditional ML implementations for the MLP model exhibited nearly identical performance metrics. Motivated by this observation, the model's architecture is enhanced to a more complex form, with the aim of testing the robustness of the TinyML conversion. The new architecture consists of 12 layers, all of which have 64 neurons with relu activation functions. The output layer is adjusted to match the number of unique classes in the dataset and employs a softmax activation function.

After modifying the architecture, a slight decline in performance is observed for the more complex model, contrasting with the unchanged accuracy in the previous simpler model, thus satisfying the initial curiosity. The findings are detailed in Table~\ref{tab:MLPPerformanceMetrics}. This slight performance drop indicates that the model's complexity may have some impact on the effectiveness of the TinyML conversion. However, this performance decrease is small and might be acceptable for certain applications.

\section{Practical Implementation and Results}

To assess the real-world applicability of TinyML, the simulation is now executed on a microcontroller. For this purpose, the ESP32 has been selected for its suitability for low-power and limited-memory environments. The popularity of this microcontroller in research is evident, as it is also utilized in other studies, such as the anomaly detection work by Antonini et al.~\cite{s23042344}. The purpose of this experiment is to validate our simulations on an actual microcontroller which allows us to compare the simulated results with real-world performance outcomes.

For the coding environment, PlatformIO, an extension in Visual Studio Code (VSCode), is used. Programming is conducted in the C language, and a client-server architecture is employed. In this setup, the ESP32 microcontroller acts as the server, while a CPU on a Linux environment, specifically Ubuntu 22.04.2 LTS, accessed via Windows Subsystem for Linux (WSL) in VSCode, acts as the client. Communication between the server (ESP32) and the client (CPU) is established using User Datagram Protocol (UDP). 

The operational process involves the client sending data samples to the ESP32 server. The server, equipped with TensorFlowLite Micro Library, performs inference on these samples. The server sends back key information to the client, including prediction time, memory usage, and the inference result. The client stores this information for future evaluation and analysis. TensorFlowLite ESP32 library by tanakamasayuki~\cite{TensorFlowLite_ESP32} is used for this deployment.

Transitioning to the practical application of this setup, due to the resource constraints of the ESP32 microcontroller, the initial MLP model is modified into a more compact version named ML\_simulated. This model employs a Sequential structure from TensorFlow's Keras library, featuring three layers with 8 neurons each, and a softmax activation function in the final layer tailored to the number of unique categories. The TensorFlow Lite library is employed to convert the model into a more compact format using default optimization strategies, resulting in a version referred to as TinyML\_simulated. This TinyML\_simulated model is then uploaded onto an ESP32 microcontroller, with the results from this deployment termed TinyML\_Microcontroller. 

Regarding the development framework, the Arduino framework is selected, with a monitor speed set at 115200 baud. The ESP32 microcontroller connects to the server through the COM3 serial port for uploading code, powering the device, and facilitating serial communication, which allows for real-time output monitoring on an external monitor.

The final stage involves compiling the code on the server side and uploading a pre-trained TFLite model to the ESP32 microcontroller. This uploading process is facilitated through Serial Peripheral Interface Flash File System (SPIFFS). After the code and model are uploaded to the microcontroller, the focus shifts to monitoring the ESP32’s connection to Wi-Fi and its reception of samples from the client. The evaluation uses the same dataset across all versions of the model, involving over 28,000 test samples. The inference times for these samples are measured, and the mean inference time is documented in Table~\ref{table:inference_time}.

\begin{table}[ht]
\centering
\caption{Comparison of Average Inference Time Per Data-sample Across Simulated and Microcontroller Deployed Models}
\begin{tabular}{|c|c|c|}
\hline
\textbf{Model} & \textbf{Average Inference Time} \\
\hline
ML\_simulated & 82.095 ms \\
\hline
TinyML\_simulated & 230.926 $\mu$s \\
\hline
TinyML\_Microcontroller & 246.825 $\mu$s \\
\hline
\end{tabular}
\label{table:inference_time}
\end{table}

\begin{table*}[ht]
\centering
\caption{Comparison of Performance Metrics Across Simulated and Microcontroller Deployed Models}
\begin{tabular}{|c|c|c|c|c|}
\hline
\textbf{Model}             & \textbf{Accuracy (\%)} & \textbf{Precision (\%)} & \textbf{Recall (\%)} & \textbf{F1 Score (\%)} \\ \hline
ML\_Simulated               & 99.9823                & 99.9827                 & 99.9823              & 99.9825                \\ \hline
TinyML\_Simulated           & 99.9823                & 99.9827                 & 99.9823              & 99.9825                \\ \hline
TinyML\_Microcontroller     & 80.8376                & 80.8365                 & 80.8376              & 80.8370                \\ \hline
\end{tabular}
\label{table:performance_metrics_ESP}
\end{table*}

The inference time of TinyML\_Microcontroller is approximately 1.07 times longer than that of TinyML\_simulated. This difference in prediction time is because of the distinct capabilities of the CPUs used in each system. The ESP32 is equipped with an Xtensa Dual-Core CPU operating at 240 MHz, which, while efficient for a microcontroller, is significantly less powerful than the Intel Core i7-9700 CPU used in the computer system. This difference in computing power directly affects how quickly the ESP32 can process data and perform inferences.

The performance metrics comparison between the ML-Simulated, TinyML-Simulated, and TinyML-Microcontroller models is presented in Table~\ref{table:performance_metrics_ESP}. The observed difference in the performance metrics of the TinyML\_Microcontroller model can be primarily attributed to the unique constraints of the ESP32 environment. While the desktop environment operates with 64-bit floating point precision, ESP32 is designed to handle 32-bit operations to reduce computational load~\cite{ESP32_TechRef}. This variance in computational precision directly influences the mathematical operations, leading to less accurate or approximated calculations in the TinyML\_Microcontroller model compared to TinyML\_Simulated. Additionally, the ESP32's design prioritizes energy efficiency and low power consumption, crucial for battery-operated or portable devices, necessitating a compromise in computational capabilities.

\section{Future Research Directions}

\subsection{Decentralized Federated Learning Approaches} In the EVCS ecosystem, the transmission of data to a CMS is essential for optimizing station performance, usage, and maintenance. Incorporating Federated Learning (FL) into this framework can significantly enhance privacy and security. FL offers a decentralized learning approach, enabling charging stations to share only model updates rather than sensitive data~\cite{liu2023recent}. This method ensures privacy, as individual data from EVCS is not transferred to a central server, allowing EVs to benefit from shared data without compromising privacy~\cite{AGRAWAL2022346}. Furthermore, FL accommodates the unique needs of different EVCS. Instead of applying a uniform model across the entire network, FL allows each sub-network to develop its own model. This approach leads to greater specificity and effectiveness for each EVCS according to their specific characteristics and demands.

\subsection{Adapting to New Threats using Online Learning} Within the context of EVCI, which this research has examined, new threats are likely to emerge over time. A static IDS, unable to learn from evolving data patterns, would be inadequate in detecting such new threats. Currently, TinyML models are generally trained on advanced computational systems and then transferred to MCUs. This process yields a static model that cannot easily learn from new data or adapt to various situations, limiting its effectiveness in dynamic IoT environments. Future research could focus on developing TinyML models capable of adjusting to new information and specific needs~\cite{ren2021tinyol}. Online learning is a potential solution to this limitation. It enables TinyML to perform on-device training on streaming data, thus maintaining the model's knowledge of previously learned threats while adapting to new ones.

\subsection{Integrating Explainable AI (XAI) with TinyML} The evolution of TinyML presents a significant opportunity to enhance the security and functionality of IDS within the EVCI domain. However, as these systems become more autonomous, ensuring their decisions are transparent and understandable becomes crucial. This is where Explainable AI (XAI) could play a pivotal role. By integrating XAI with TinyML, the critical need for transparency in automated threat detection and response mechanisms can be addressed. XAI facilitates the transition from black-box models to more interpretable white-box models~\cite{kabir2022explainable}. Moreover, adding XAI to TinyML not only enhances transparency but also aids in refining models. Future research could benefit from explainability in TinyML, which can potentially make systems such as EVCI more reliable.

\subsection{Trustworthy TinyML} Trustworthiness in AI is a multifaceted concept encompassing robustness, generalization, explainability, transparency, reproducibility, fairness, privacy preservation, and accountability. These dimensions collectively contribute to making AI systems reliable, ethical, and socially beneficial. Achieving trust in AI requires attention to its entire process, from data collection and model creation to deployment and ongoing monitoring. Such an all-encompassing strategy is vital for aligning AI with ethical norms and societal expectations, essential for improving trust among users and stakeholders~\cite{li2023trustworthy}. As research on trustworthy AI progresses, the need to incorporate trustworthiness into TinyML becomes increasingly evident. Given that TinyML involves compressing AI models to fit on small, power-constrained devices, this process may impact the models' accuracy and performance. Consequently, there is a stronger need to explore how the principles of trustworthiness—robustness, generalization, explainability, and others—can be maintained or adapted within the TinyML context.

\section{Conclusion}

While traditional ML models have paved the way for advancements in cybersecurity, their high computational and energy demands often limit their applicability in real-time and resource-constrained environments. In this paper, a comprehensive review of TinyML is conducted, highlighting its significant advantages over traditional ML models in the context of cybersecurity. By bringing computations closer to edge devices, TinyML reduces latency, enhances privacy, energy efficiency, memory usage, and cost-effectiveness. The TinyML application in IDS, specifically in the context of EVCI, is explored as the case study. Comparison between the TinyML model and traditional model reveals that the former significantly outperforms in terms of reduced time and memory usage. A practical implementation on an ESP32 is also executed to validate these findings. Ultimately, the crucial role that TinyML can play in enhancing the security and privacy of real-time applications is underscored by this research.

\bibliographystyle{ieeetr}
\bibliography{references.bib}

\end{document}